\documentclass[aps,pra,reprint,twocolumn,superscriptaddress,floatfix,longbibliography, nofootinbib]{revtex4-2}
\usepackage{microtype}
\usepackage{graphicx, amsfonts, amsthm, xr}
\usepackage{array}
\usepackage{epsfig,amsmath, amssymb, color, dsfont, upgreek, physics}
\usepackage{mathrsfs}
\usepackage{mathtools}
\usepackage{bbold}
\usepackage{comment}
\usepackage{braket}
\usepackage{lipsum}
\usepackage{float}
\usepackage{soul}
\usepackage{bm}

\usepackage[bookmarks=true,colorlinks,linkcolor=OrangeRed,urlcolor=NavyBlue,citecolor=RoyalBlue]{hyperref}
\usepackage[capitalise]{cleveref}
\Crefname{figure}{Fig.}{Figs.}
\Crefname{section}{Sec.}{Secs.}
\usepackage[dvipsnames]{xcolor}
\usepackage{orcidlink}
\usepackage[inline]{enumitem}
\setlist[enumerate]{label=(\roman*)}

\usepackage{parskip}
\newcommand{\tdot}{.\@}

\newcommand{\idest}{i.e\tdot}

\definecolor{darkgreen}{rgb}{0,0.5,0}
\definecolor{redred}{HTML}{D53E4F}
\newcommand{\danger}[1]{{\color{redred} #1}}

\newcommand{\gtext}[1]{{\color{darkgreen} #1}}
\definecolor{mygold}{rgb}{0.93,0.69,0.13}

\definecolor{mypurple}{rgb}{0.49,0.18,0.56}

\newcommand{\rla}{\danger{r}}
\newcommand{\gla}{\gtext{g}}

\newcommand{\mindex}[1]{m_{\mathrm{#1}}}
\newcommand{\mL}{\mindex{L}}
\newcommand{\mR}{\mindex{R}}

\usepackage{stackengine, scalerel}

\def\cmp{\mathbin{\ThisStyle{\ensurestackMath{\abovebaseline[-\dimexpr1.1pt+0.55\LMpt]{%
  \stackunder[-\dimexpr1pt+2.5\LMpt]{\color{darkgreen}\SavedStyle-}{%
  \color{red}\SavedStyle+}}}}}}

\newcommand{\jonehalf}{\hspace{-0.7pt}{\frac12}\hspace{-0.7pt}}

\let\ket\relax 
\DeclarePairedDelimiter\ket{\lvert}{\rangle}

\newcommand\blfootnote[1]{
  \begingroup
  \renewcommand\thefootnote{}\footnote{#1}
  \addtocounter{footnote}{-1}
  \endgroup
}

\newcommand{\orcidgiuseppeC}{\orcidlink{0000-0002-5749-2224}}
\newcommand{\orcidgiuseppeM}{\orcidlink{0000-0002-7280-445X}}
\newcommand{\orcidpietro}{\orcidlink{0000-0001-5279-7064}}
\newcommand{\orcidsimone}{\orcidlink{0000-0002-8882-2169}}
\newcommand{\orcidgiovanni}{\orcidlink{0000-0002-9073-8978}}
\newcommand{\orcidmarco}{\orcidlink{0000-0002-4544-3513}}
\newcommand{\orciddarvin}{\orcidlink{0000-0001-8805-3761}}
\newcommand{\orcidjad}{\orcidlink{0000-0002-0659-7990}}

\newcommand{\DFA}{\affiliation{Dipartimento di Fisica e Astronomia ``G. Galilei'', Università di Padova, I-35131 Padova, Italy.}}
\newcommand{\PQTC}{\affiliation{Padua Quantum Technologies Research Center, Università degli Studi di Padova}}
\newcommand{\INFNPD}{\affiliation{Istituto Nazionale di Fisica Nucleare (INFN), Sezione di Padova, I-35131 Padova, Italy.}}
\newcommand{\bari}{\affiliation{Dipartimento di Fisica, Università di Bari, I-70126 Bari, Italy.}}
\newcommand{\INFNBA}{\affiliation{Istituto Nazionale di Fisica Nucleare (INFN), Sezione di Bari, I-70125 Bari, Italy.}}
\newcommand{\LMU}{\affiliation{Department of Physics and Arnold Sommerfeld Center for Theoretical Physics (ASC), Ludwig Maximilian University of Munich, 80333 Munich, Germany}}
\newcommand{\MPQ}{\affiliation{Max Planck Institute of Quantum Optics, 85748 Garching, Germany}}
\newcommand{\MCQST}{\affiliation{Munich Center for Quantum Science and Technology (MCQST), 80799 Munich, Germany}}
\newcommand{\Berlin}{\affiliation{Dahlem Center for Complex Quantum Systems, Free University of Berlin, 14195 Berlin, Germany}}

\graphicspath{{figures_prl/}}
\begin{document}

\title{Quantum Many-Body Scarring in a Non-Abelian Lattice Gauge Theory}

\author{Giuseppe Calaj\'o\orcidgiuseppeC *} \INFNPD
\author{Giovanni Cataldi\orcidgiovanni *} \INFNPD \DFA \PQTC
\author{Marco Rigobello\orcidmarco} \INFNPD \DFA \PQTC
\author{Darvin Wanisch\orciddarvin} \INFNPD \DFA \PQTC
\author{Giuseppe Magnifico\orcidgiuseppeM} \bari \INFNBA
\author{Pietro Silvi\orcidpietro} \INFNPD \DFA \PQTC
\author{Simone Montangero\orcidsimone} \INFNPD \DFA \PQTC
\author{Jad C.~Halimeh \orcidjad}\email{jad.halimeh@lmu.de} \LMU \MPQ \MCQST \Berlin
\blfootnote{* These authors contributed equally to this work.}
\date{\today}

\begin{abstract}
    Quantum many-body scarring (QMBS) is an intriguing mechanism of weak ergodicity breaking that has recently spurred significant attention. Particularly prominent in Abelian lattice gauge theories (LGTs), an open question is whether QMBS nontrivially arises in non-Abelian LGTs. 
    Here, we present evidence of robust QMBS in a non-Abelian $\mathrm{SU}(2)$ LGT with dynamical matter.
    Starting in product states that require little experimental overhead, we show that prominent QMBS arises for certain quenches, facilitated through meson and baryon-antibaryon excitations, highlighting its non-Abelian nature. The uncovered scarred dynamics manifests as long-lived coherent oscillations in experimentally accessible local observables as well as prominent revivals in the state fidelity. 
    Our findings bring QMBS to the realm of non-Abelian LGTs, highlighting the intimate connection 
    between scarring and gauge symmetry, and are amenable for observation in a recently proposed trapped-ion qudit quantum computer.
\end{abstract}

\maketitle

The nature of equilibration in an isolated interacting quantum many-body system is a fundamental question in physics \cite{Reimann2008}. Whereas open quantum systems are expected to thermalize due to exchanging energy with a bath \cite{Reichental2018}, the conditions under which an isolated quantum many-body system thermalizes are stipulated within the Eigenstate Thermalization Hypothesis (ETH) \cite{Deutsch1991,Srednicki1994,Rigol_review, Deutsch_review}.
An \textit{ergodic} Hamiltonian is expected to lead to \textit{quantum thermalization} for generic initial states, where the information of the initial state is locally erased \cite{Rigol_2008}.

In recent years, a new paradigm of weak ergodicity breaking has emerged that violates the ETH. The typical scenario consists, in certain ergodic Hamiltonians, 
in a polynomial (in system size) number of special non-thermal \textit{scar} eigenstates, that are roughly equally spaced in energy over the whole spectrum \cite{Turner2018,BernevigEnt,Schecter2019}, and that exhibit anomalously low bipartite entanglement entropy \cite{lin2018exact}.
In some specific models, scars states can also occur as a continuous band of low entanglement-entropy states \cite{PhysRevResearch.5.043208} or as isolated states, such as in the case of the AKLT model \cite{PhysRevB.98.235156}.

Upon initializing the system in an initial state with a high overlap with these scar eigenstates, the subsequent quench dynamics give rise to \textit{quantum many-body scarring} (QMBS), which manifests as prominent oscillations in local observables and persistent revivals in the (local and global) Loschmidt fidelity that last longer than the typical system's timescales, 
thereby circumventing expected thermalization \cite{Bernien2017, Moudgalya2018, Zhao2020, Jepsen2021, Serbyn2020, Moudgalya_review, Chandran_review}. 
QMBS has received a lot of attention since its initial discovery in \cite{Bernien2017}, and has since been the subject of many quantum simulation experiments \cite{Bluvstein2021, Bluvstein2022quantum, Su2022, Zhang2023Many-body, Dong2023Disorder}.

After the discovery of QMBS, it has been shown that the effective model quantum-simulated in \cite{Bernien2017} is a spin-$1/2$ $\mathrm{U(1)}$ lattice gauge theory (LGT), where the electric field is represented by a spin-$1/2$ $z$-operator \cite{Surace2020}. The robustness of QMBS also depends on the stability of the underlying gauge symmetry in this model \cite{Halimeh2022robust}. QMBS also arises in spin-$S$ $\mathrm{U}(1)$ LGTs \cite{Desaules2022weak, Desaules2022prominent}, in LGTs with other Abelian gauge groups \cite{Iadecola2020quantum, aramthottil2022scar, desaules2024massassisted}, and in $2+1$D LGTs \cite{Banerjee2021, biswas2022scars, ebner2024entanglement, Sau2024,osborne2024quantum,budde2024quantum}. 
However, all of these works involve Abelian gauge groups except the recent Ref.~\cite{ebner2024entanglement}, which considers a pure (without dynamical matter) $\mathrm{SU(2)}$ LGT.
Looking at the bipartite entanglement entropy, they identify a few scar-eigenstate candidates in a certain parameter regime. However, no experimentally feasible scarred initial states have been proposed, and the type of scarring exhibited vanishes beyond the crudest truncation of the electric field basis.

Given the intimate connection between QMBS and LGTs, and the current large effort to quantum simulate the latter \cite{Dalmonte_review,Pasquans_review,Zohar_review,Alexeev_review,aidelsburger2021cold,zohar2021quantum,klco2021standard,Bauer_review,dimeglio2023quantum,halimeh2023coldatom,cheng2024emergent}, it is important to investigate the origin of QMBS in connection to gauge symmetry by exploring its possible occurrence in non-Abelian LGTs with dynamical matter. In particular, to faithfully investigate a high-energy context of QMBS, including dynamical matter is crucial, since nature hosts dynamical matter fields. In this work, we provide exact diagonalization (ED) and matrix product state (MPS) results that showcase robust QMBS dynamics in a $1+1$D  non-Abelian $\mathrm{SU}(2)$ LGT with dynamical matter starting in simple initial product states. We show how the QMBS dynamics involve state transfer through meson and baryon--anti-baryon bare states, highlighting the non-Abelian nature of this scarred dynamics. We further present preliminary results suggesting that such scar dynamics persists in $2+1$ dimensions for parameter regimes similar to those discussed in the $1+1$D scenario.

\begin{figure*}
    \includegraphics{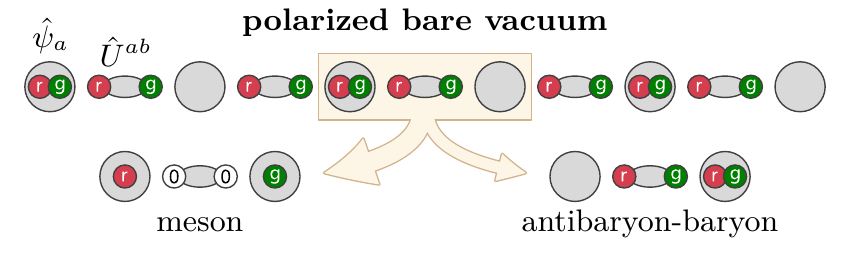}\hfill\includegraphics{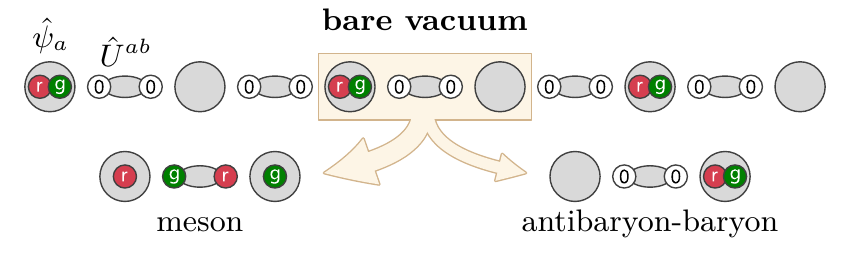}
    \includegraphics{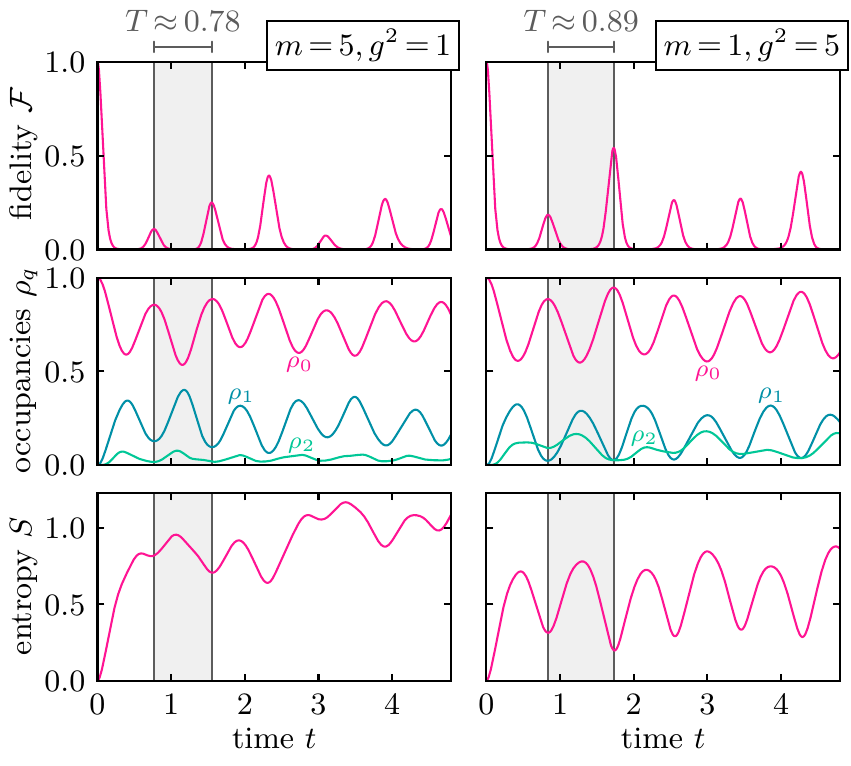}\hfill\includegraphics{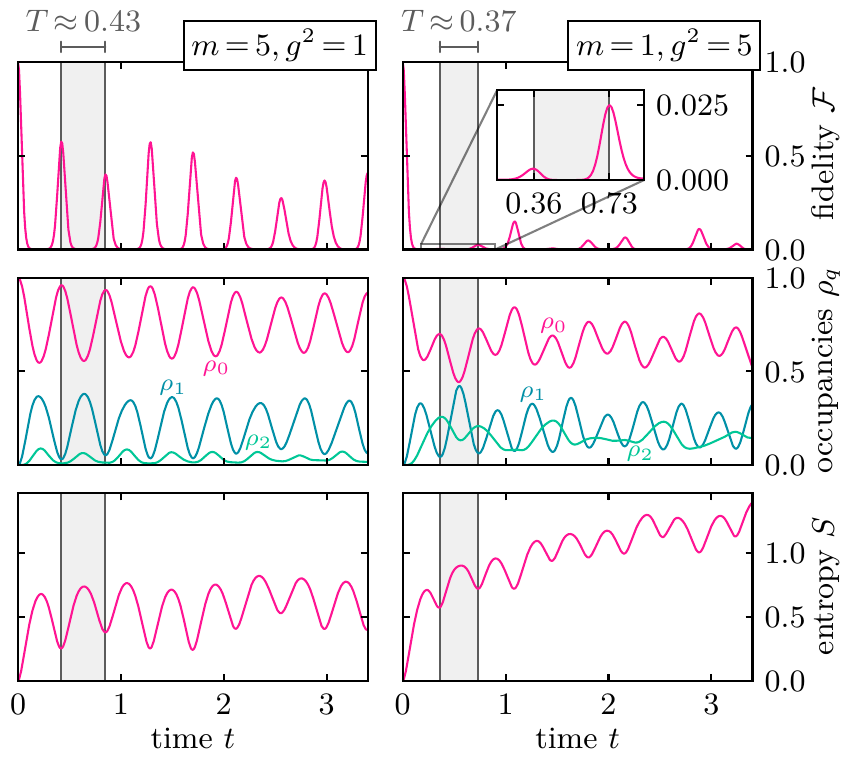}
    \caption{Many-body scarring dynamics of the polarized bare vacuum (left) and the bare vacuum (right) initial states.
        \emph{Top:} cartoons depicting the classical configurations with leading contributions to the dynamics; with circles and ellipses denoting matter sites and gauge links respectively.
        \emph{Bottom:} return fidelity, average quark occupancy, and bipartite entanglement entropy as a function of time,
        for different mass and coupling regimes, reported in Figure.
        MPS simulation with $N=30$ sites in open boundary conditions (OBC), maximum bond dimension $D_{\rm max}=350$, and truncation tolerance $\rm tol = 10^{-7}$.
    }\label{fig_dynamics}
\end{figure*}

\section{Model} As a prototypical model to probe non-ergodic dynamics in non-Abelian LGTs, we consider a $1+1$D matter-coupled hardcore-gluon \cite{Cataldi2024SimulatingSUYangMills, rigobello2023hadrons, Calajo2024DigitalQuantumSimulation, Silvi2019Tensor, Zohar2019Removing} $\mathrm{SU}(2)$ LGT.
The dynamics is governed by the Kogut-Susskind Hamiltonian, on an $N$-site lattice with spacing $a_0$ ($c,\hbar=1$) \cite{Kogut1975}:
\begin{multline}\label{eq_H}
    \hat{H}_0 =
    \frac{1}{2a_0} \sum_{n\vphantom{a,b\in\{\rla,\gla\}}} \sum_{a,b\in\{\rla,\gla\}}
    \qty[-i\hat\psi^{\dagger}_{na}\hat U^{ab}_{n,n+1}\hat\psi_{n+1,b}+{\rm H.c.} ] \\
    + m_0\sum_{n,a} (-1)^{n} \hat\psi^{\dagger}_{na}\hat\psi_{na}
    + \frac{a_0 g_0^2}{2} \sum_n \hat E^2_{n,n+1}
    \,.
\end{multline}
The model describes a matter quark field of mass $m_0$, living on lattice sites $n$,
coupled, with strength $g_0$, to a truncated SU(2) gauge field, defined on lattice links | see \cref{fig_dynamics}.
The quark field is a staggered fermion doublet $\hat\psi_{na}$, obeying anticommutation relations
$\acomm*{\hat\psi_{na}}{\hat\psi^{\dagger}_{n'b}}{=} \delta_{n,n'} \delta_{a,b}$ and $\acomm*{\hat\psi_{na}}{\hat\psi_{n'b}}{=}0$
\cite{susskind1977lattice}.
A basis for matter sites is
\begin{math}
    \{
    {\ket{0}},\,
    {\ket{\rla} = \hat \psi^{\dagger}_{\rla}\ket{0}},\,
    {\ket{\gla} = \hat \psi^{\dagger}_{\gla}\ket{0}},\,
    {\ket{2}    = \hat \psi^{\dagger}_{\rla} \hat \psi^{\dagger}_{\gla}\ket{0}}
    \}
\end{math}.
Gauge link states can be expanded in the the chromoelectric basis basis $\ket{j, \mL, \mR}$,
 where $j \in \mathbb{N}/2$ indicates the spin irreducible representations and 
$\mR,\mL\in\{-j,\ldots,+j\}$ label  the states within the  spin shell $j$.
On this basis
 the link energy density operator is diagonal and coincides with the quadratic Casimir, $\hat{E}^2\ket{j, \mL, \mR}=j(j\mathop+1)\ket{j, \mL, \mR}$ \cite{Zohar2015}.
Adopting a hardcore-gluon truncation, we restrict $j$ to $\{0,1/2\}$ for most of the following analysis, except in Fig. \ref{fig_j1}, where we demonstrate the persistence of scar dynamics at higher gauge field truncation. This approximation
 keeps only basis states reachable from the singlet $\ket{00}$ via at most a single application of the parallel transporter $\hat U^{ab}$:
\begin{math}
    \{\ket{00},\ket{\rla\rla},\ket{\gla\gla},\ket{\gla\rla},\ket{\rla\gla}\}
\end{math},
where $\rla,\gla$ are short-hands for $\pm1/2$ and $j$ was left implicit.

Non-Abelian Gauss law mandates that physical states are local gauge singlets,
\begin{math}
    \hat{G}^\nu_n \ket{\Psi_{\text{phys}}} \equiv 0, \; \forall n,\,\nu\in\{x,y,z\}
\end{math},
where $\hat{G}^\nu_n$ are the generators of local rotations at site $n$.
In numerical simulations, we enforce this constraint using a dressed site approach, which yields a defermionized qudit model with a $6$-dimensional local basis \cite{Calajo2024DigitalQuantumSimulation}. Its derivation is detailed in App. \ref{AppA}.
In the qudit model, it is convenient to remove the explicit dependence on $a_0$ by rescaling the energy, mass, and coupling in terms of the dimensionless quantities
$(\hat{H},m,g^2) = 4 \sqrt{2} a_0 (\hat{H}_{0},m_0,3a_0g_0^2/16)$.
Values reported in the Figures refer to these rescaled quantities.

\section{Non-Abelian scarred dynamics}
To investigate the presence of non-ergodic behavior in this model, we consider the Schrödinger time evolution of different initial states $\ket{\Psi(t\mathop=0)}$, across various parameter regimes.
The time evolution is performed on a chain of $N=30$ sites  with open boundary conditions (OBC) via MPS methods.
The details of the simulation are reported in App. \ref{AppC}.
As initial configuration, we first consider the \emph{polarized bare vacuum} (PV), which consists of the matter sites in the bare vacuum configuration, while the gauge sites in the excited electric field state,
\begin{equation}
    \ket{\Psi_{\text{PV}}}{=}
    \ket{\dots2020\dots}_{\mathrm{m}}
    \ket*{\cdots 
        \frac{\rla\gla-\gla\rla}{\sqrt{2}}
        \cdots
        \frac{\rla\gla-\gla\rla}{\sqrt{2}}               
        \cdots
    }_{\!\!\mathrm{g}}
    ,
    \label{eq_pol_vacuum_state}
\end{equation}
where $\ket{\:\cdot\:}_\mathrm{m}$ and $\ket{\:\cdot\:}_\mathrm{g}$ refer to the matter and gauge subsystems, respectively.
Here $(\mR)_{n-1,n}$ and $(\mL)_{n,n+1}$ around each bulk site $n$ are valence-bonded in a singlet.
In the qudit formulation of the model, this state becomes a product state (see App. \ref{AppA}).

We identify two regimes avoiding thermalization:
the moderately large-mass one, $m> (g^2,1)$, and the moderately large-coupling one, $g^2> (m,1)$.
For each regime, we compute the return fidelity between the initial state and the evolved state, $\mathcal{F}(t)=\abs{\braket{\Psi(t)|\Psi(0)}}^2$, shown in \cref{fig_dynamics}.
After an initial complete system relaxation, we observe persistent revivals, signaling a clear deviation from the expected ergodic behavior.
The revivals occur approximately periodically, with the period indicated in the figure for both the considered regimes. 

The observed dynamics can be understood as a process of persistent particle pair creation out of an initial false vacuum~\cite{Banerjee2012, Kuehn2014, Magnifico2020realtimedynamics}.
To see this, we define the site occupancy operator,
\begin{equation}
    \hat Q_n = (-1)^n \sum_a \left[\hat\psi^{\dagger}_{na}\hat\psi_{na}-\frac{1-(-1)^n}{2}\right]
    \,,
\end{equation}
which counts the number of quarks (on even sites) or antiquarks (on odd sites).
Then, we compute the average occupancy $\rho_{q}(t)$ of each $\hat Q_n$ eigenvalue (single site quark occupancy, $q\in\{0,1,2\}$),
projecting on the corresponding eigensubspace of $\hat Q_n$ and averaging over all sites $n$:
\begin{equation}
    \rho_{q}(t) = \frac{1}{N} \sum_{n} \ev{\delta_{\hat Q_n, q}}{\Psi(t)}\,.
\end{equation}
They oscillate around average values that do not coincide with the ones predicted by thermal ensembles, confirming deviations from ergodic behavior, as detailed in App. ~\ref{AppB}.
In the moderately large-mass regime, most of the excitations are meson-like (quark and antiquark pairs adjacent to an excited shared gauge link). 
 In contrast, in the moderately large-coupling regime, the formation of baryon-like (quark pair) and antibaryon-like (antiquark pair) excitations becomes more probable, as indicated by the enhanced double occupancy compared to the large-mass regime. 
 Importantly, compared to the case of large values of either mass or coupling, where particle pair creation is a rare event that confines the system's dynamics close to its initial state, for moderately large values of these parameters, multiple pairs can be created, allowing the system to escape its initial configuration completely. This is evidenced by the fidelity dropping to zero in \cref{fig_dynamics}. 

Finally, the last row of \cref{fig_dynamics} shows the evolution of the bipartite entanglement entropy defined as $S=-{\rm Tr}[\hat \varrho_A\log \hat\varrho_A]$, where $A$ and $B$ indicate the two halves of the chain and $\hat \varrho_A={\rm Tr}_B[\hat \varrho_{AB}]$ is the reduced density operator of subsystem A.
In both cases, we observe a slow growth of the entanglement entropy after an initial fast increase up to the first fidelity revival peak. The oscillations on top of the growth are driven by the successive fidelity revivals, which indeed have the same period.

In addition to the polarized bare vacuum initial state, we also observe similar scarred-like dynamics for the \textit{bare vacuum} (V) state: the ground state of the Hamiltonian in \cref{eq_H} without the hopping term, made of alternated doubly occupied and empty sites, and trivial links,
\begin{equation}
    \ket{\Psi_{\text{V}}(t=0)}=\ket{\dots2020\dots}_{\mathrm{m}} \ket{\dots0000\dots}_{\mathrm{g}}
    \,.
    \label{eq_vacuum_state}
\end{equation}
The resulting non-ergodic dynamics is also shown in \cref{fig_dynamics}.
Compared to the polarized vacuum case, the oscillation's period in the fidelity revivals and occupancy is smaller.
It can be explained by the many-body spectrum as discussed in the following section.
The moderately large coupling regime is characterized by a faster entanglement growth, consistently with a corresponding lower return fidelity.
Although in the regimes considered this product state is energetically close to the true ground state of the model, nontrivial physics drives the observed dynamics. 
This is evident from the significant oscillations observed in the average occupancy, which lead the system significantly away from its initial configuration.

\begin{figure*}
    \includegraphics{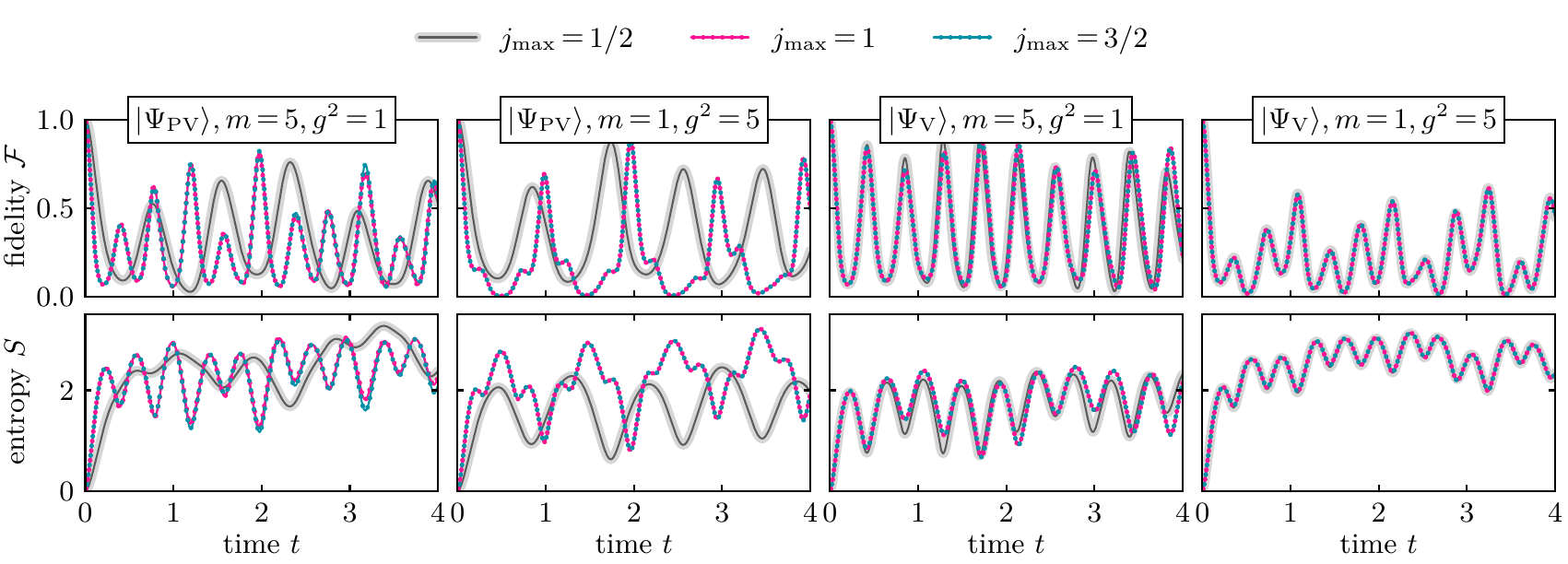}
    \caption{\textit{Scarring at higher gauge truncation.} Many body scarring dynamics for the polarized
        bare vacuum and the bare vacuum for the truncated SU(2) YM LGT at $j_{\max}=1$ (pink line) and $j_{\max}=3/2$ (cyan line) in comparison with the corresponding one at $j_{\max}=1/2$ (grey line).
        Each column reports the fidelity and bipartite entanglement entropy as a function of time for the same two cases $(m,g)$ considered in \cref{fig_dynamics}.
        The simulations are obtained via ED for $N=8$ sites in PBC}.
    \label{fig_j1}
\end{figure*}

For both initial states, non-ergodic dynamics persist at longer times and remain robust with increasing system size, as we discuss in \cref{AppB}. 
The scar dynamics observed in Fig. \ref{fig_dynamics} is not an artifact of the adopted gauge field truncation, as demonstrated in \cref{fig_j1}, where we consider the  same quenches and parameter regimes as in Fig. \ref{fig_dynamics}, this time including representations up to $j=1$ and $j=3/2$.
As the figure shows, the  $j=1$ and $j=3/2$ truncations give negligible differences in the return fidelity and entanglement entropy, confirming the validity of the link truncation in the studied regimes. 
This is expected for large coupling or mass values, where the link field's truncation is well-justified, as the hopping term acts as a perturbation relative to these dominant energy scales.

Finally, we stress that the observed dynamical behavior is not observed for other initial state configurations or different coupling-mass regimes (see \ref{AppB}), suggesting that the non-ergodicity originates directly from many-body scars in the spectrum. 
In the following, we demonstrate it by analyzing the full many-body spectrum of the model.

\begin{figure*}
    \includegraphics{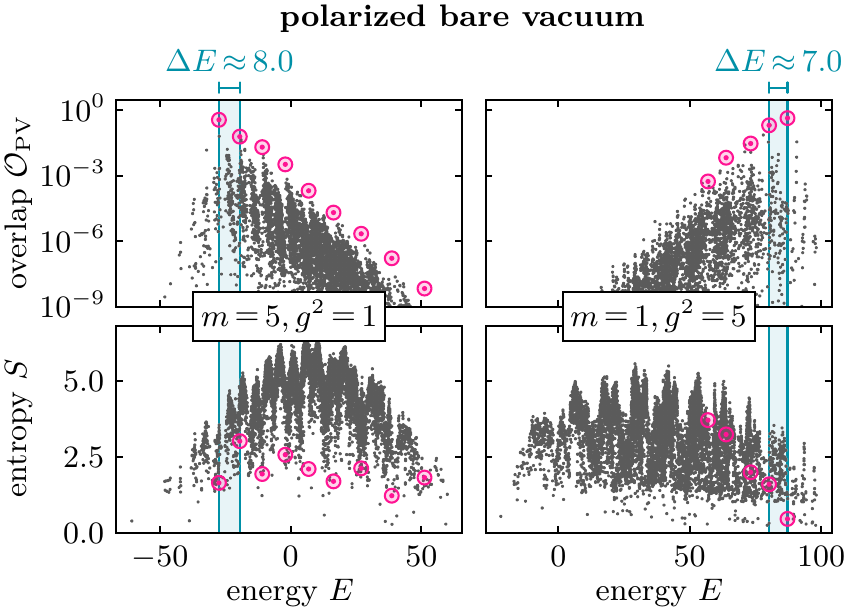}\hfill\includegraphics{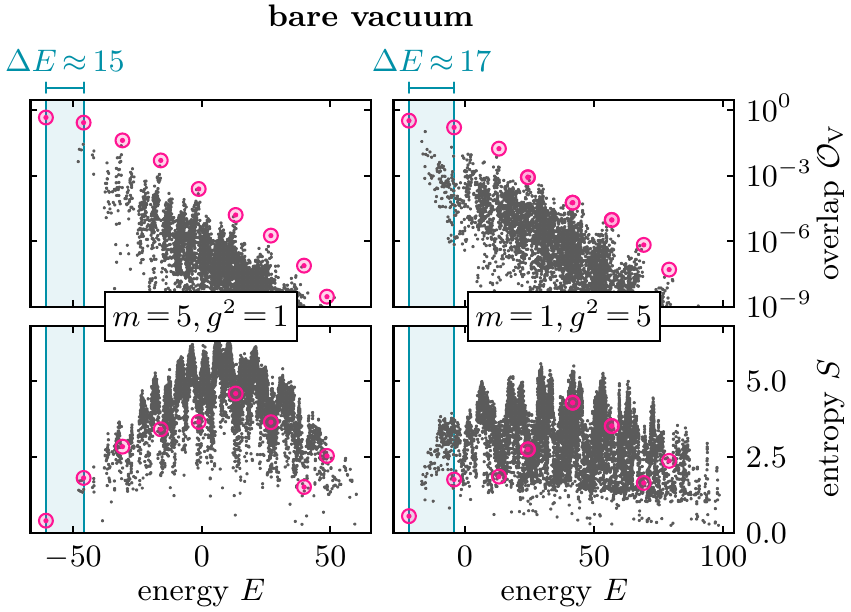}
    \caption{Spectrum analysis for a lattice chain of $N\mathop=10$ sites in OBC, in the two parameter regimes of \cref{fig_dynamics}.
    Red circles highlight the scar states.
    \emph{First row:} the overlap of the many-body spectrum with the polarized bare vacuum (left) and the bare vacuum (right);
    only states with $\mathcal{O}_{\mathrm{(P)V}}>10^{-9}$ are included.
    \emph{Second row:} bipartite entanglement entropy of each eigenstate.
    }
    \label{fig_tower}
\end{figure*}

\section{Tower of scar states}
To determine whether the observed non-ergodic dynamics originates from quantum many-body scars, we perform ED with OBC up to $N=10$ sites and, for each eigenstate in the many-body spectrum $\{\ket{\Phi_s}\}$,
we compute the overlap
\begin{math}
    \mathcal{O}_{\mathrm{PV/V}}
    = \abs*{\braket{\Psi_{\mathrm{PV/V}}|\Phi_s}}^2
\end{math}
and the bipartite entanglement entropy $S$ 
with the two considered initial states (polarized, PV, and unpolarized bare vacua, V).
These quantities are plotted in \cref{fig_tower}, for both regimes where we observe revivals in the return fidelity.
For each candidate QMBS regime, we find a `tower of scar states' (highlighted by red circles) characterized by a high overlap with the initial product states and entanglement low enough (compared to the rest of the many-body spectrum) to give rise to scarring behavior.
In each case, the energy gap between the scar states in the tower, $\Delta E$ (reported in \cref{fig_tower}), approximately matches the frequency of the revivals observed in the return fidelity from \cref{fig_dynamics}, $\Delta E \approx 2\pi/T$.
The correspondence between spectral and dynamical features leads us to conclude that QMBS is indeed the underlying mechanism behind the non-ergodic dynamics shown in \cref{fig_dynamics}.

\section{Ergodic vs non-ergodic dynamics}
\begin{figure}
    \includegraphics{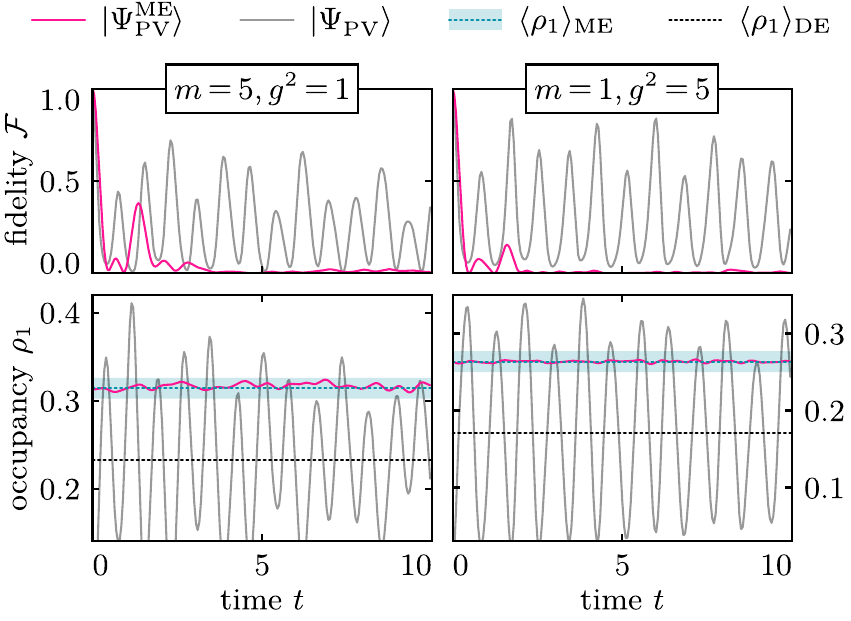}
    \caption{
        Dynamics of the \textit{microcanonical} state in \cref{eq_thermal_state} for the same parameter regimes of \cref{fig_dynamics}.
        Gray curves reproduce the polarized bare vacuum dynamics, for comparison.
        The blue horizontal lines reproduce the thermal $\rho_1$ occupancy, obtained via microcanonical ensemble (ME).
        The black ones show the diagonal ensemble (DE) results for the long-time average of $\rho_1$ on $\ket{\Psi_{\rm PV}(t)}$, highlighting its lack of thermalization.
        ED results for $N=10$ sites in PBC.
    }
    \label{fig_thermal_dyn}
\end{figure}

To confirm that the observed non-ergodic dynamics is indeed due to scarring, we should observe thermal relaxation for other initial states at the same energy.
To verify this, we focus on the case of the polarized vacuum state and evolve the \textit{microcanonical} ensemble (ME) state constructed as a uniform coherent superposition of all eigenstates lying within a small energy window around the quench energy $E_{\rm PV}{=} \ev*{\hat{H}}{\Psi_{\rm PV}}$:
\begin{equation}
    \label{eq_thermal_state}
    \ket{\Psi^{\rm ME}_{\rm PV}} = \frac{1}{\sqrt{N_{E_{\rm PV},\delta E}}} \sum\nolimits_{s,\: |E_s-E_{\rm PV}|<\delta E} \ket{\Phi_s},
\end{equation}
where $\delta E{=}\sqrt{\ev*{\hat{H}^2}{\Psi_{\rm PV}} - E_{\rm PV}^2}$ defines the energy shell $[E_{\rm PV}{-}\delta E, E_{\rm PV}{+}\delta E]$,
and $N_{E_{\rm PV},\delta E}$ is the number of eigenstates in this energy shell \cite{Rigol_2008}.
Figure~\ref{fig_thermal_dyn} compares the evolution of this state to that of $\ket{\Psi_{\rm PV}}$, in the identified scarred regimes   under periodic boundary conditions (PBC).
As the plot illustrates, the \textit{microcanonical} state exhibits ergodic behavior:
return fidelities relax toward zero (up to minor oscillations attributable to the finite lattice size) and the single occupancy remains close to its microcanonical ensemble average $\ev*{\rho_1}_{\rm ME}$ \cite{Rigol_2008}.
In contrast, for the polarized vacuum, $\rho_1$ widely oscillates around its long-time average $\ev*{\rho_1}_{\rm DE}$, whose value is predicted by the diagonal ensemble (DE) and largely deviates from the thermal result (see App. \ref{AppB}).
We thus conclude that the observed non-ergodic dynamics strictly depends on the initial conditions and escapes an otherwise thermal behavior through the occurrence of scar states in the many-body spectrum.

\section{Many-body scarring in $2+1$D}\label{Sec2D}
In this section we perform a preliminary numerical analysis to investigate the presence of many-body scars in a non-Abelian SU(2) LGT with dynamical matter in $2+1$D. 
We employ the model described in Ref. \cite{Cataldi2024SimulatingSUYangMills} where SU(2)-color staggered fermions $\hat\psi_{\bm{n},a}
$ of mass $m_0$ are placed in a $N=L_x\times L_y$ lattice $\Lambda$ with lattice spacing $a_0$. Here sites and link are respectively identified by the couple $(\bm{n},\bm{\mu})$, where $\bm{n}=(n_x,n_y)$ indicates lattice sites, while $\bm{\mu}$ can be one of the two positive lattice unit vectors $\bm{\mu}_x=(1,0)$ or $\bm{\mu}_y=(0,1)$.
The Hamiltonian of the model reads ($c,\hbar=1)$:
\begin{equation}
\begin{split}
    \label{eq_H2D}
    \hat{H}_0=&
    \frac{1}{2a_0}
    \sum_{\bm{n}}\sum_{a,b}
    \big[-i\hat\psi^{\dagger}_{\bm{n},a}
    \hat U^{ab}_{\bm{n},\bm{n}+\bm{\mu}_x}\hat\psi_{\bm{n}+\bm{\mu}_x,b}\\
    &\qquad\quad -(-1)^{n_x+n_y}\hat\psi^{\dagger}_{\bm{n},a}
    \hat U^{ab}_{\bm{n},\bm{n}+\bm{\mu}_x}\hat\psi_{\bm{n}+\bm{\mu}_x,b}+{\rm H.c.}\big] \\
    &+ m_0\sum_{\bm{n},a} (-1)^{n_x+n_y} \hat\psi^{\dagger}_{\bm{n},a}
    \hat\psi_{\bm{n},a}\\
    &+ \frac{g_0^2}{2a_0} \sum_{\bm{n},k}\hat E^2_{\bm{n},\bm{n}+\bm{\mu}_{k}}
    -\frac{1}{2g_0^2a_0}\sum_{\square\in\Lambda}\Tr(U_{\square}+U^{\dag}_{\square}).
\end{split}
\end{equation}
The last term in Eq. \eqref{eq_H2D} represents the SU(2)-magnetic energy density approximated by the smallest Wilson loops \cite{Cataldi2024SimulatingSUYangMills} defined as
\begin{equation}
    U_{\square} = \sum_{a,b,c,d}
    U_{\bm{n}, \bm{\mu}_{x}}^{ab}
    U_{\bm{n}+\bm{\mu}_{x},\bm{\mu}_{y}}^{bc}
    U_{\bm{n}+\bm{\mu}_{y},\bm{\mu}_{x}}^{cd\dagger}
    U_{\bm{n},\bm{\mu}_{y}}^{da\dagger},
\end{equation}
with $\bm{\mu}_{x}$ and $\bm{\mu}_{y}$ spanning the plaquette's xy-plane.
Similarly to the $1+1$D case, we consider the smallest nontrivial truncation of the link degree of freedom employing the $(0{\otimes} 0){\oplus}(1/2{\otimes}1/2)$ representation of the SU(2) gauge field.

\begin{figure}
    \includegraphics{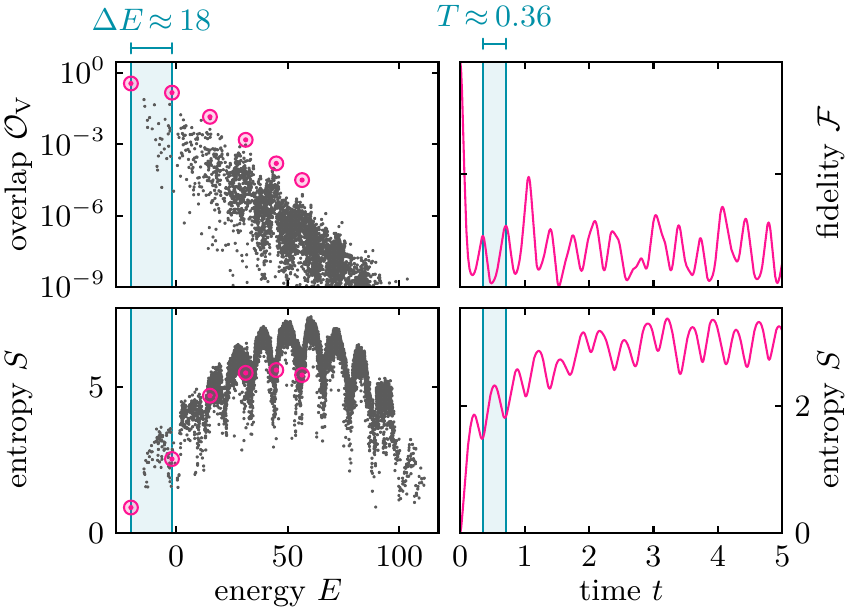}%
    \caption{\textit{Quantum many-body scarring in $2+1$D.} (left) Many-body spectrum analysis for a $4\times 2$ lattice ladder in OBC at $m=5$ and $g^{2}=1$.
    Red circles highlight the scar eigenstates with high overlap with the bare vacuum state (top) and low bipartite entanglement entropy (bottom).
    (right) Many-body scarring dynamics of the bare vacuum initial state regarding the return fidelity (top), and the bipartite entanglement entropy as a function of time (bottom). The oscillating period $T$ is compatible with the energy spacing $\Delta E$ between scar states in the spectrum.
    }
    \label{fig_scars2D}
\end{figure}

For this model, we perform ED simulations on a ladder of $N=4\times 2$ sites with OBC, exploring the same mass and coupling regimes as in the $1+1$D case. 
In \cref{fig_scars2D}, we focus on the 2D analog of the bare vacuum state considered in \cref{eq_vacuum_state}, characterized by a staggered double occupancy of lattice sites and inactive gauge links.
Despite the small system size, we observe the emergence of a `tower of scar states' (highlighted by red circles), which further exhibit low entanglement entropy. 
Correspondingly, the dynamics displays revivals in fidelity and an oscillating slowly increasing entanglement entropy. Even if such oscillating behavior could be attributed to  finite-size effects,
the dynamical patterns are characterized by an oscillating period $T$ comparable to the energy separation $\Delta E$ between the sequent tower of states in the spectrum, similarly as observed  for the $1+1$D case.
Comparison with the pure gauge results of Ref. \cite{ebner2024entanglement} suggests that dynamical matter may play a crucial role in many-body scarring beyond one spatial dimension, where the magnetic term effect plays an important role. 
We leave the investigation of larger system sizes to future work.

\section{Summary and outlook
}
In this letter, we presented strong numerical evidence for the occurrence of quantum many-body scars in a non-Abelian $\mathrm{SU}(2)$ lattice gauge theory with dynamical matter.
In particular, starting from simple experimentally friendly vacuum product states, scarred dynamics manifests as persistent oscillations in local observables and periodic revivals in the state fidelity. This dynamics arises from a state transfer through meson and baryon-antibaryon bare states, which allow the system to escape otherwise expected thermalization.
Even though the results were obtained for a non-Abelian lattice gauge theory model with the gauge-link truncated at the lowest nontrivial irreducible representations, 
we demonstrated that in the regimes where we observe scars — moderately large values of coupling or mass — the truncation of the link field is well-justified and the observed scar dynamics persist even with higher gauge-link truncations.
This is similar to what has recently been observed for Abelian models \cite{Desaules2022prominent}, suggesting that quantum many-body scarring survives in the Kogut-Susskind limit of the SU(2) model under consideration.
These findings demonstrate that the strong connection between QMBS and LGTs persists even for non-Abelian gauge groups, raising the intriguing question of whether such behavior represents an intrinsic feature of more fundamental theories such as quantum chromodynamics.

Our work opens the door to exciting future directions: In the App. \ref{AppB}, we present numerical evidence of scar eigenstates with very low bipartite entanglement entropy throughout the entire spectrum even in regimes where both $m$ and $g$ are quite small. 
It would be interesting to find simple product states showing large overlap with these eigenstates and to see if these states survive at higher truncations of the link, being the truncated model, not a good approximation of Eq.~\eqref{eq_H} at small $g$.
Additionally, similarly to the PXP model \cite{Daniel2023}, it would be interesting to map the scarring phase diagram of this model to see how driving protocols can enhance non-Abelian scarring \cite{Hudomal2022} and to perform a deeper analysis on the underlying algebraic structures as done  in Ref.~\cite{Moudgalya2023ExhaustiveCharacterizationQuantum}.

Another venue to pursue is exploring the robustness of the non-Abelian scarring uncovered in this work in two spatial dimensions, as it has been recently done for Abelian gauge theories~\cite{osborne2024quantum,Budde2024}. 
In~\ref{Sec2D}, we performed a preliminary analysis of this scenario by solving an SU(2) truncated $2+1$D model~\cite{Cataldi2024SimulatingSUYangMills} with dynamical matter on a ladder of $N=4\times 2$ sites. Our results suggest that many-body scars may also occur in this setting for similar parameter regimes to those discussed in this paper. We leave a deeper analysis on larger system sizes for future work.
Furthermore, scarring is known to give rise to complex periodicities in the corresponding dynamical quantum phase transitions of Abelian LGTs~\cite{VanDamme2022dynamical,VanDamme2023anatomy}; it would be then interesting to see if this carries over to the non-Abelian case.

Finally, we stress that, as addressed in Ref.~\cite{Calajo2024DigitalQuantumSimulation},
the qudit formulation of the proposed model is well-suited for conducting a digital quantum simulation on a recently demonstrated trapped-ion qudit quantum processor~\cite{ringbauer2022universal}. This opens up the intriguing possibility to experimentally observe this exotic non-ergodic dynamics in the near future.

\section{Acknowledgments}
The authors thank Anthony Ciavarella, Jean-Yves Desaules, and Paolo Stornati for valuable discussions. They acknowledge financial support from:
the European Union via QuantERA2017 project QuantHEP,
via QuantERA2021 project T-NiSQ,
via the Quantum Technology Flagship project PASQuanS2,
and via the NextGenerationEU project CN00000013 - Italian Research Center on HPC, Big Data and Quantum Computing (ICSC);
the Italian Ministry of University and Research (MUR) via PRIN2022-PNRR project TANQU,
via Progetti Dipartimenti di Eccellenza project Frontiere Quantistiche (FQ) and project Quantum Sensing and Modelling for One-Health (QuaSiModO);
the Max Planck Society, the Emmy Noether Programme of the German Research Foundation (DFG) via  grant no.~HA 8206/1-1, the Deutsche Forschungsgemeinschaft (DFG, German Research Foundation) under Germany’s Excellence Strategy – EXC-2111 – 390814868, and the European Research Council (ERC) under the European Union’s Horizon Europe research and innovation program (Grant Agreement No.~101165667)—ERC Starting Grant QuSiGauge;
the WCRI-Quantum Computing and Simulation Center (QCSC) of
Padova University;
the University of Bari through the 2023-UNBACLE-0244025 grant
the Istituto Nazionale di Fisica Nucleare (INFN) via project NPQCD and project QUANTUM.
The authors also acknowledge computational resources by Cloud Veneto and the ITensors Library for the MPS calculations~\cite{ITensor,ITensor-r0.3} as well as the \textrm{ed-lgt} Library for Exact Diagonalization \cite{Cataldi2024}.
This work is part of the Quantum Computing for High-Energy Physics (QC4HEP) working group.

\appendix

\section{Hardcore-gluon qudit model}\label{AppA}

In order to map the Kogut-Susskind SU(2) lattice Yang-Mills Hamiltonian of \cref{eq_H} to the qudit model used in numerical simulations,
we:
\begin{enumerate*}
    \item perform a hardcore-gluon truncation of the gauge field;
    \item build the gauge-singlet local dressed basis; and
    \item gauge-defermionize.
\end{enumerate*}
The procedure is conveniently formulated by decomposing matter and gauge local Hilbert spaces in SU(2) irreducible representations (irreps).
This preliminary step is already built in the local bases introduced in the main text:
the irrep basis $\ket{j, \mL, \mR}$ for gauge links,
and the Fock basis
\begin{math}
    \{\ket{q}\}
    =
    \{
    {\ket{0}},\,
    {\ket{\rla},\ket{\gla} = \hat\psi^{\dagger}_{\rla,\gla}\ket{0}},\,
    {\ket{2}    = \hat\psi^{\dagger}_{\rla} \hat\psi^{\dagger}_{\gla}\ket{0}}
    \}
\end{math}
for matter sites.
Matter basis states are associated with the following spin labels $(j,m)$:
\begin{equation}
    \ket{0}               \leftrightarrow (0,0)
    \,\quad
    \ket{\rla},\ket{\gla} \leftrightarrow (0,\pm1/2)
    \,\quad
    \ket{2}               \leftrightarrow (0,0)
    \,.
\end{equation}
Irrep decomposition specifies how local gauge rotations act on a site $n$ and its neighboring links.
Infinitesimal rotations are generated by
\begin{math}
    \hat{\mathbf{G}}_{n} = \hat{\mathbf{R}}_{n-1,n} + \hat{\mathbf{Q}}_{n} + \hat{\mathbf{L}}_{n,n+1}
\end{math},
where
\begin{math}
    \hat{Q}^{\nu}_{n} =
    \sum_{ab}
    \hat\psi^{\dagger}_{na}
    S^{({\scriptscriptstyle\jonehalf})\nu\!}_{ab}
    \hat\psi_{nb}
\end{math}
rotates the quark field at $n$,
while $\hat{\mathbf{R}}_{n-1,n}$ and $\hat{\mathbf{L}}_{n,n+1}$
account for the transformation of the gauge links at its left and its right respectively \cite{Zohar_review}:
\begin{subequations}\label{eq_LR_rishon_rotations}
    \begin{align}
        \label{eq_L_rishon_rotation}
        \langle j' \mL' \mR' |
        \hat L^{\nu}
        | j \mL \mR \rangle & =
        \delta_{j,j'}
        S^{(j)\nu}_{\mL',\mL} \delta_{\mR',\mR}
        \,,
        \\
        \label{eq_R_rishon_rotation}
        \langle j' \mL' \mR' |
        \hat R^{\nu}
        | j \mL \mR \rangle & =
        \delta_{j,j'}
        \delta_{\mL',\mL} S^{(j)\nu}_{\mR',\mR}
        \,;
    \end{align}
\end{subequations}
$S^{(j)\nu}$ being the spin-$j$ $\mathfrak{su}(2)$ matrices, with $\nu\in\{x,y,z\}$.
The chromoelectric energy operator is just their square
\begin{math}
    \hat E^2 =  \hat{\mathbf{R}}^2 = \hat{\mathbf{L}}^2
\end{math},
\idest{} the quadratic Casimir \cite{Zohar_review}:
\begin{align}
    \hat E^2
    | j \mL \mR \rangle & =
    j(j+1)
    | j \mL \mR \rangle
    \,.
\end{align}
In principle, the gauge field could occupy arbitrarily high spin shells; however, in the hardcore-gluon model, we restrict to $j\in\{0,1/2\}$, corresponding to a cutoff $\norm*{\hat E^2} \leq 3/4$ on the chromoelectric field operator spectrum ($\norm{\:\cdot\:}$ denotes the matrix norm).
Recovering the untruncated physics is not crucial for the present study.
Nevertheless, we stress that hardcore gluons give a good approximation of the untruncated low energy physics in the strong coupling limit $g_0\gg1$, where the gauge field energy term dominates the Hamiltonian.

To enforce the physical state condition, \idest{}, non-Abelian Gauss law,
\begin{math}
    \hat{\mathbf{G}}_{n}  \ket{\Psi_{\text{phys}}} \equiv 0
\end{math},
we observe that $\hat{\mathbf{L}}$ and $\hat{\mathbf{R}}$ from \cref{eq_LR_rishon_rotations} act nontrivially only on the $\mL$ and $\mR$ index respectively.
This suggests \cite{Silvi_2014} to factorize each gauge link in a pair of new rishon degrees of freedom, living at its edges.
Upon fusing each site with its adjacent rishons, we forge a composite site where Gauss law becomes an internal constraint \cite{rigobello2023hadrons}.
We label the basis states of each rishon as
\begin{math}
    \ket{0}
    = \ket{j\,{=}\,0,m\,{=}\,0}
\end{math}
and
\begin{math}
    \ket{\rla},
    \ket{\gla}
    = \ket{j\,{=}\,\frac{1}{2},m\,{=}\,\pm\frac{1}{2}}
\end{math}.
The left and right rishons on each link are then constrained to be in the same spin shell $j$ by restricting to the even parity sector of a local $\mathbb{Z}_2$ symmetry, whose action reads
\begin{equation}\label{eq_link_symmetry}
    \hat P = +\op{0} - (\op{\rla} + \op{\gla})
    \,.
\end{equation}
In this way, the original physical link states are selected.
Solving Gauss law in the composite site amounts to finding the local singlets via Clebsh-Gordan decomposition,
which provides the following 6-dimensional local dressed basis,
\begin{math}
    \{
    \ket{\alpha} =
    \ket{\mR^{\alpha}({n-1,n}),\,q^{\alpha}(n),\,\mL^{\alpha}({n,n+1})}
    \}_{\alpha=1}^6
\end{math} \cite{Calajo2024DigitalQuantumSimulation}:
\begin{equation}\label{eq_dressed basis}
    \begin{aligned}
        \ket{1} & =\ket{0,0,0}                                          , &
        \ket{2} & =\frac{\ket{\rla,0,\gla}-\ket{\gla,0,\rla}}{\sqrt{2}} ,   \\
        \ket{3} & =\frac{\ket{\gla,\rla,0}-\ket{\rla,\gla,0}}{\sqrt{2}} , &
        \ket{4} & =\frac{\ket{0,\rla,\gla}-\ket{0,\gla,\rla}}{\sqrt{2}} ,   \\
        \ket{5} & =\ket{0,2,0}                                          , &
        \ket{6} & =\frac{\ket{\rla,2,\gla}-\ket{\gla,2,\rla}}{\sqrt{2}} .
    \end{aligned}
\end{equation}

In the irrep basis for gauge link, the parallel transporter is given in terms of Clebsh-Gordan coefficients \cite{Zohar2015}:
\begin{equation*}\label{eq_parallel_transporter}
    \langle j' \mL' \mR' |
    \hat U^{ab}
    | j \mL \mR \rangle =
    \sqrt{\frac{2j+1}{2j^{\mathrlap{\prime}}+1}}\:
    \overline{C^{j,\mL}_{\jonehalf,a;\,j',\mL'}}
    C^{j',\mR'}_{\jonehalf,b;\,j,\mR}
    .
\end{equation*}
we realize it in terms of rishon operators as
\begin{equation}\label{eq_parallel_transporter_rishons}
    \hat U^{ab}_{n,n+1} \to \frac{1}{\sqrt{2}}
    \hat\zeta^{\mathrm{(L)} a}_{n,n+1}
    (\hat\zeta^{\mathrm{(R)} b}_{n,n+1})^\dagger
    \,,
\end{equation}
with
\begin{equation}
    \hat{\zeta}^{\rla} = \ketbra{0}{\rla} + \ketbra{\gla}{0}
    \,,\quad
    \hat{\zeta}^{\gla} = \ketbra{0}{\gla} - \ketbra{\rla}{0}
    \,.
\end{equation}
Observe that the right-hand side of \cref{eq_parallel_transporter_rishons} preserves the link parity, as desired.
Conversely, a single rishon operator always inverts the local parity:
\begin{math}
    \acomm*{\hat P}{\hat\zeta^a} = 0
\end{math}.
The existence of a $\mathbb{Z}_2$-grading in the rishon spaces allows to gauge-defermionize the model \cite{Cataldi2024SimulatingSUYangMills}:
we take $\hat\zeta^a$ to be fermionic and introduce the bosonic operators,
\begin{equation}
    \hat Q_n^{\mathrm{(L,R)}} = \sum_a (\hat\zeta^{\mathrm{(L,R)} a}_{n,n+1})^\dagger \hat\psi_{n,a}
    \,,\quad
    \hat M = \sum_a \hat\psi_{n,a}^\dagger\hat\psi_{n,a}
    \,.
\end{equation}
For potential quantum simulation implementations, it is convenient to rewrite these solely in terms of Hermitian operators, introducing \cite{Calajo2024DigitalQuantumSimulation}
\begin{equation}
    \begin{aligned}
        \hat{A}^{(1)} & \mathbin=         \hat Q^{\mathrm{(L)}} \mathop+ \hat{Q}^{\mathrm{(L)}\dagger}               \,, &
        \hat{B}^{(1)} & \mathbin=         \hat Q^{\mathrm{(R)}} \mathop+ \hat{Q}^{\mathrm{(R)}\dagger}               \,,   \\
        \hat{A}^{(2)} & \mathbin= i \big[ \hat Q^{\mathrm{(L)}} \mathop- \hat{Q}^{\mathrm{(L)}\dagger} \big] \,,         &
        \hat{B}^{(2)} & \mathbin= i \big[ \hat Q^{\mathrm{(R)}} \mathop- \hat{Q}^{\mathrm{(R)}\dagger} \big] \,.
    \end{aligned}
\end{equation}
Finally, we map \cref{eq_H} to the following 6-dimensional qudit Hamiltonian \cite{Calajo2024DigitalQuantumSimulation}:
\begin{equation}
    \label{eq_Heff}
    \hat H
    = \sum_{\mathclap{j\in\{1,2\}}} \; \sum_n \hat A^{(j)}_n \hat B^{(j)}_{n+1}
    + m (-1)^{n} \hat M_n
    + g^2 \sum_n \hat C_n
    \,,
\end{equation}
where we rescaled energy to absorb the hopping prefactor,
$\hat H = 4 \sqrt{2} a_0 \hat H_{0}$, and defined
the dimensionless couplings
$m = 4 \sqrt{2} a_0  m_0$ and
$g^2 = \frac{3 \sqrt{2}}{4} a_0 g_0^2$.
The qubit operators appearing in \cref{eq_Heff} are reported below \cite{Calajo2024DigitalQuantumSimulation}:
\begin{subequations}
    \begin{align}
        \hat Q^{\mathrm{(L)}} & = \sqrt{2}\ketbra{1}{4}+\ketbra{2}{3}+\ketbra{3}{6}+\sqrt{2}\ketbra{4}{5} \,, \\
        \hat Q^{\mathrm{(R)}} & = \sqrt{2}\ketbra{1}{3}+\ketbra{2}{4}+\sqrt{2}\ketbra{3}{5}+\ketbra{4}{6} \,, \\
        \hat{M}               & = \ketbra{3}{3}+\ketbra{4}{4}+2\ketbra{5}{5}+2\ketbra{6}{6}               \,, \\
        \hat{C}               & = 2\ketbra{2}{2}+\ketbra{3}{3}+\ketbra{4}{4}+2\ketbra{6}{6}               \,.
    \end{align}
\end{subequations}
Finally, in the dressed qudit basis given in \cref{eq_dressed basis}  the initial states considered in the main text read respectively 
\begin{equation}
 |\Psi_{\text{PV}}\rangle=| 6\rangle| 2\rangle...| 6\rangle| 2\rangle\,,  
\end{equation}
for the polarized bare vacuum and 
\begin{equation}
 |\Psi_{\text{V}}\rangle=| 5\rangle| 1\rangle...| 5\rangle| 1\rangle\,,
\end{equation}
for the bare vacuum.

\subsection{Scars at higher gauge link truncation}\label{spin_1_scars}

The rishon decomposition of $\hat{U}^{ab}$ proposed in \cref{eq_parallel_transporter_rishons} can be generalized to arbitrary truncation of the maximum allowed spin shell $j_{\text{max}}$, although at a (manageable) added cost.
Starting from a given spin shell $j$, we have to separately account for the action when both rishons are increased to shell $j+\frac{1}{2}$, and both are decreased to shell $j-\frac{1}{2}$.
We can then decompose $\hat{U}^{ab}$ as follows \cite{Cataldi2024SimulatingSUYangMills}:
\begin{equation}
    \hat{U}^{ab}_{\vb{j}, \vb{j}+\vb*{\mu}}=\hat{\zeta}_{A,\vb{j},\vb*{\mu}}^{a} \hat{\zeta}_{B,\vb{j}+\vb*{\mu},-\vb*{\mu}}^{b\dagger}+\hat{\zeta}_{B,\vb{j},\vb*{\mu}}^{a} \hat{\zeta}_{A,\vb{j}+\vb*{\mu},-\vb*{\mu}}^{b\dagger},
    \label{eq_SU2_U_definition}
\end{equation}
where the two $\zeta$-rishon species, A and B, act respectively as raising and lowering the spin shell of the SU(2) gauge irreducible representation. Interestingly, they are related to each other as:
\begin{align}
    \hat{\zeta}_{A}^{a} & = i \sigma^{y}_{a,b} \hat{\zeta}_{B}^{b\dagger}, &
    \hat{\zeta}_{A}^ {a\dagger} = i \sigma^{y}_{a,b} \hat{\zeta}_{B}^{b}.
\end{align}
We can then rewrite \cref{eq_SU2_U_definition} just in terms of one species, e.g. B.
Dropping the index, i.e. $\hat{\zeta}_{B}^{a}=\hat{\zeta}^{a}$, it holds:
\begin{equation}
    \hat{U}^{ab}_{\vb{j}, \vb{j}+\vb*{\mu}}=
    i \sigma^{y}_{a,c} \hat{\zeta}_{\vb{j},\vb*{\mu}}^{c\dagger} \hat{\zeta}_{\vb{j}+\vb*{\mu},-\vb*{\mu}}^{b\dagger}
    + i \sigma^{y}_{b,c} \hat{\zeta}_{\vb{j},\vb*{\mu}}^{a} \hat{\zeta}_{\vb{j}+\vb*{\mu},-\vb*{\mu}}^{c}.
\end{equation}
For a chosen truncation $j_{\text{max}}$ of the SU(2) irreducible representation, $\zeta$-rishons are defined as follows \cite{Cataldi2024SimulatingSUYangMills}:
\begin{equation}
    \hat{\zeta}^{\gla(\rla)}{=}\qty[\sum_{j=0}^{j_{\text{max}}-\frac{1}{2}}\sum_{m=-j}^{j}\chi(j,m,\gla(\rla)) \ket{j,m}\bra{j+\mbox{$\frac{1}{2}$},m{\cmp}\mbox{$\frac{1}{2}$}}]_F,
    \label{eq_SU2_general_rishon}
\end{equation}
where the function $\chi(j,m,\alpha)$ reads
\begin{equation}
    \chi\qty(j,m,\gla(\rla))=\sqrt{\frac{j\cmp m+1}{\sqrt{(2j+1)(2j+2)}}}.
\end{equation}
For instance, truncating the gauge link to the shell $j=1$, we obtain a local Hilbert qudit basis of $10$ states while truncating to $j=3/2$ the local basis is made of $14$ states.

\section{Ergodic vs non-ergodic behavior}\label{AppB}
In this section, we provide further evidence that the non-ergodic dynamics discussed in the main text are indeed attributable to the occurrence of many-body scars. Specifically, we first demonstrate how, for other parameter regimes an ergodic behavior is observed as predicted by thermal ensembles. Secondly, we illustrate how scars-induced revivals persist at larger system sizes and over longer timescales.
\subsection{Thermal ensembles}
The long-time average of a generic observable $\hat O$ resulting from the unitary dynamics of a given initial state $\ket{\Psi(0)}=\sum_s C_s\ket{\Phi_s}$ expressed in the eigenbasis of the Hamiltonian \eqref{eq_H} is well described by the expectation value predicted by the \emph{diagonal} ensemble \cite{Rigol_2008}:
\begin{equation}
    \begin{split}
        \bar{O}&\equiv\lim_{T\to\infty} \frac{1}{T}\int_0^T dt \braket{\Psi(t)|\hat O|\Psi(t)}                   \\
        &=       \lim_{T\to\infty} \frac{1}{T}\int_0^T dt
        \sum_{s,p} C^{*}_{s}C_{p}\braket{\Phi_{s}|e^{it\hat{H}}\hat Oe^{-it\hat{H}}|\Phi_{p}}                                                       \\
        &= \sum_{s,p} C^{*}_{s}C_{p}\braket{\Phi_{s}|\hat O|\Phi_{p}}\lim_{T\to\infty} \frac{1}{T}\int_0^T e^{i(E_{s}-E_{p})t} dt \\
        &=       \sum_{s}\abs{C_s}^2 \braket{\Phi_s|\hat O|\Phi_s}=
        \Tr \{\hat{O}\hat{\varrho}_{\rm DE}\}\equiv\braket{\hat{O}}_{\rm DE},
    \end{split}
\end{equation}
where $\hat{\varrho}_{\rm DE}=\sum_{s}\abs{C_s}^2 \ketbra{\Phi_s}$ is the diagonal ensemble.
Then, if the expectation value of the observable $\braket{\hat{O}(t)}$ approaches long-time average $\braket{\hat{O}}_{\rm DE}$ over long times, the system \emph{equilibrates}.
Rather, \emph{thermalization} is achieved when the steady state coincides with the prediction of the microcanonical ensemble whose thermal average reads
\begin{equation}
    \braket{\hat O}_{\rm ME}={\rm Tr}\{\hat O  \hat{\varrho}_{\rm ME} \}\,,
\end{equation}
where the microcanonical ensemble is defined as a superposition within the energy shell $[E_\text{q}-\delta E , E_\text{q}+\delta E ]$, with the quench energy $E_\text{q}=\bra{\Psi(0)}\hat{H}\ket{\Psi(0)}$,  containing $N_{E,\delta E }$ eigenstates:
\begin{equation}
    \label{eq_microcanonical_state}
    \hat{\varrho}_{\rm ME}=\frac{1}{N_{E,\delta E }}\sum_{\substack{s;\\|E_s-E_\text{q}|<\delta E}}\ketbra{\Phi_s}{\Phi_s}\,.
\end{equation}

\subsection{Ergodic dynamics}
\begin{figure}
    \includegraphics{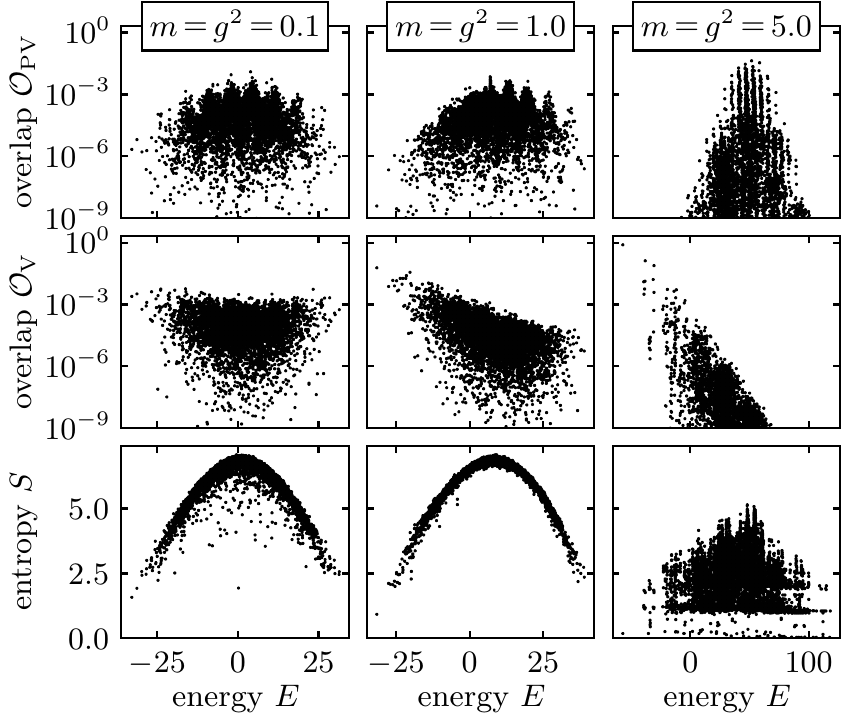}%
    \caption{\textit{Many-body spectrum.} Spectrum analysis for a lattice chain of $N = 10$ sites with open boundary conditions in a few paradigmatic regimes (corresponding to different columns) where we observe ergodic dynamics.
    The first and second rows show the overlap of the many-body spectrum with the bare vacuum and polarized bare vacuum state, respectively.
    Only states with $\mathcal{O}_{\mathrm{PV}}>10^{-9}$ are included in the plot.
    In the third row, we display the bipartite entanglement entropy of all eigenstates in the energy spectrum as a function of their respective eigenvalues.}
    \label{fig_various_spectra}
\end{figure}

We have shown that the two considered initial product states polarized bare vacuum (PV) and bare vacuum (V), escape thermalization for the two regimes $(m=5, g^2=1)$ and $(m=1, g^2=5)$.
To confirm the scarring nature of the observed non-ergodic dynamics we investigate the absence of scars states in the many-body spectrum for other parameter regimes.
In \cref{fig_various_spectra}, we look at the overlap of the full many-body spectrum with the two considered initial states PV and V and the corresponding bipartite entanglement entropy for different values of $g^2=m$.

\begin{figure}
    \includegraphics{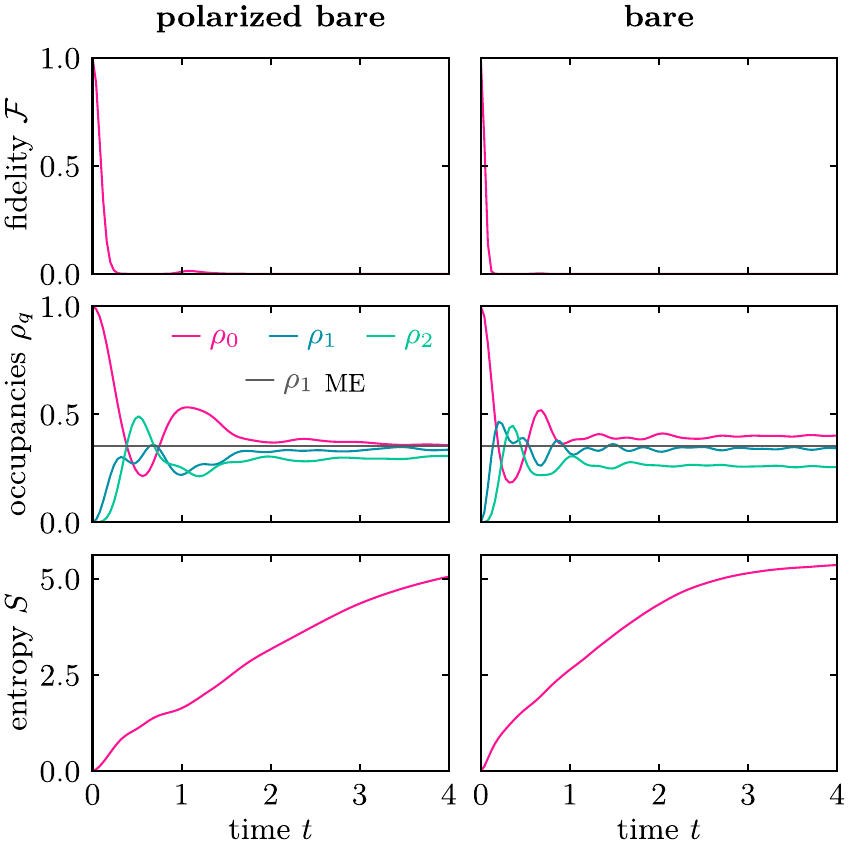}
    \caption{\textit{Ergodic dynamics.} First to third row: return fidelity, average quark occupancy, and bipartite entanglement entropy as a function of time for the two considered initial states, polarized bare vacuum (PV), and bare vacuum (V) at $m{=}g^2{=}1$.
    The horizontal lines in the occupancy panels indicate the thermal expectation value of the single particle occupancy density
    obtained through ED of a lattice of $N=10$ with PBC for the microcanonical ensemble.
    The simulation was performed with MPS for $N=20$ sites by setting OBC, maximum bond dimension $D_{\rm max}=350$, and truncation tolerance  $\rm tol=10^{-7}$.}
    \label{fig_dyn_gm1}
\end{figure}
We do not observe a clear emergence of a tower of states (rather than in \cref{fig_tower}), and thus, we expect ergodic dynamics for the two considered states.
This is also confirmed by the time evolution of the two initial states for $g^2=m=1$ displayed in \cref{fig_dyn_gm1}.
In this case, the fidelity does not exhibit persistent revivals, and the entanglement entropy shows rapid growth compared to the previously observed scarred dynamics.
Moreover, as expected for thermalization, the occupancy relaxes to a steady value compatible with the microcanonical ensemble averages.
A similar thermal behavior $g^2=m=0.1$ is observed for small values of mass and coupling when starting from the bare and polarized vacuum initial states. As shown in Fig.~\ref{fig_various_spectra}, the many-body spectrum in this regime contains states with extremely low entanglement entropy that do not overlap with these initial states. We leave further investigations into scarring dynamics for this and other candidates initial states to future work.

\begin{figure}
    \includegraphics{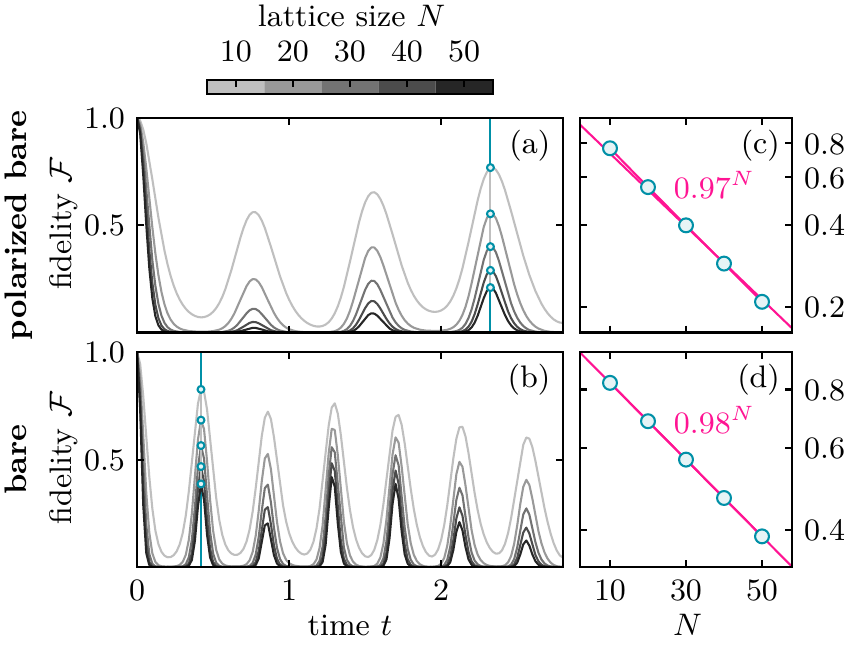}%
    \caption{\textit{Finite-size scaling.} Return fidelity during the evolution of the polarized bare vacuum (a) and bare vacuum (b) initial states, for $m=5$, $g^2=1$,  open boundary conditions, and various systems sizes $N$.
        The side plots (c) and (d) show the finite-size scaling of the highest fidelity peak, highlighted by a blue line in (a) and (b).
        Notice the log $y$-scale.
        The pink line is an exponential fit $\mathcal{F}_N=f^N$, and the extrapolated peak fidelity per site $f$ is reported in the Figure.
        Simulations performed in OBC  with maximum MPS bond dimension $D_{\rm max}=350$, and truncation tolerance  $\rm tol = 10^{-7}$.}
    \label{fig_FvsN}
\end{figure}

\subsection{Scaling with the system size}
Many-body scars represent a polynomial (in number) subset of special non-thermal eigenstates within the full many-body spectrum.
This implies that, especially in light of the well-known orthogonality catastrophe \cite{gebert2014anderson,PhysRevLett.100.080601}, the overlap with the initially considered product states decreases as the system size increases.
To investigate this behavior, in \cref{fig_FvsN}(a), we plot the return fidelity between the evolved state and two initial states discussed in the main text: the polarized bare vacuum (PV) and the bare vacuum (V).
Exploiting the slow entanglement growth in the many-body scars regime, we considered system sizes ranging from $N=10$ to $N=50$.
Although the fidelity peaks decrease with increasing $N$ as expected, pronounced revivals are still visible even for large chains with $N=50$ sites.
The peaks become sharper, with the fidelity minima approaching zero, accordingly with the expected revivals induced by scars.
The maximum amplitude of these peaks follows an exponential scaling in the system size, $\mathcal{F}_N=f^N$, with the extrapolated peak fidelity per site being $f=0.97$ and $f=0.98$ for the PV and V states, respectively. This shows that the fidelity per site is independent from the system size demonstrating the robustness of the observed scar dynamics even at large system sizes.

\subsection{Long-time dynamics}
\begin{figure*}
    \includegraphics{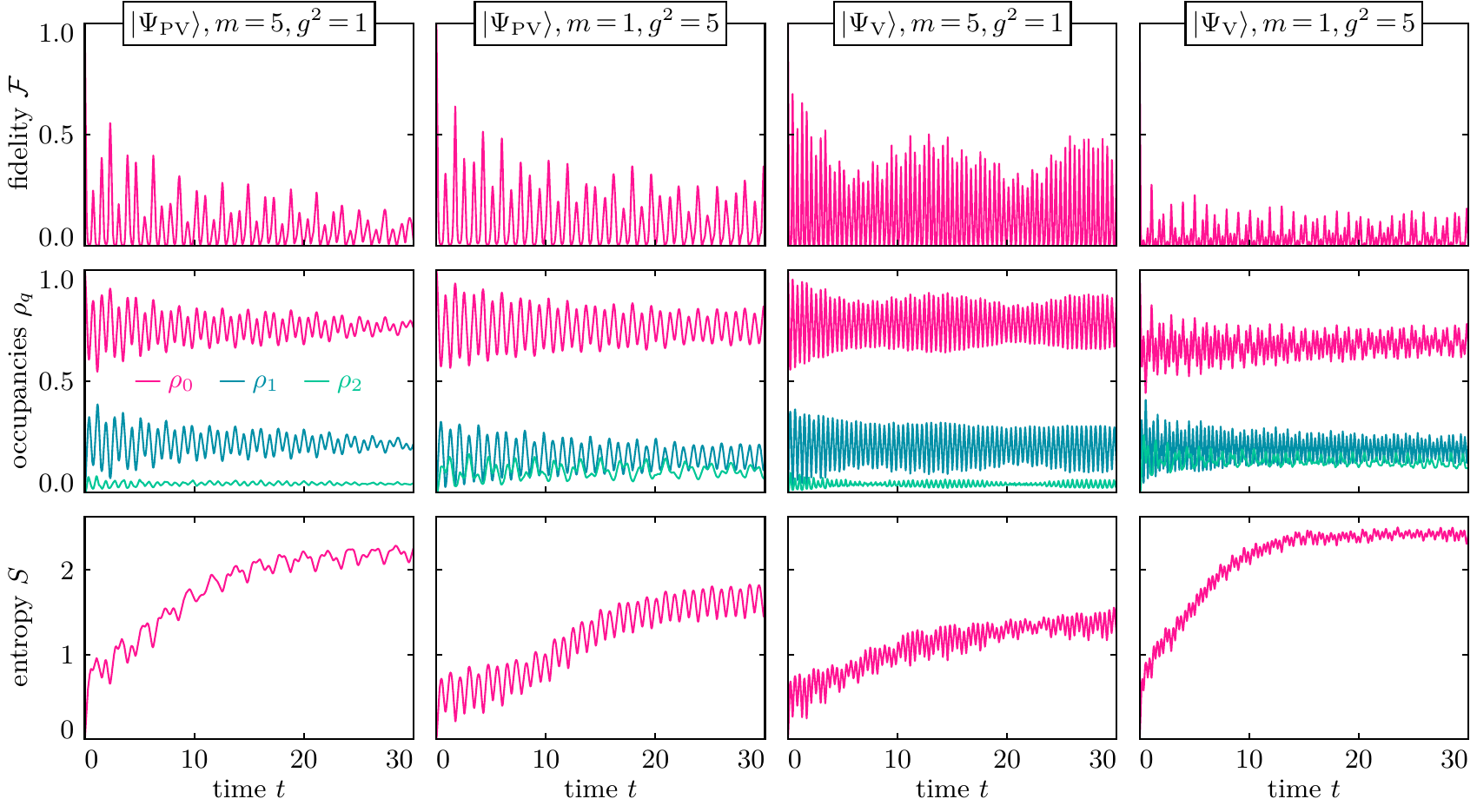}%
    \caption{\textit{Long time dynamics.} First to third row: return fidelity, average quark occupancy, and bipartite entanglement entropy as a function of time for the two considered initial states, polarized bare vacuum, and bare vacuum.
        We considered different values of mass and coupling as indicated in the Figure. The simulation was performed with MPS for $N=30$ sites by setting  OBC, maximum bond dimension $D_{\rm max}=350$, and truncation tolerance  $\rm tol = 10^{-7}$.}
    \label{fig_Flongt}
\end{figure*}

To illustrate how the observed non-ergodic dynamics persist over long times, we conducted simulations for an extended duration.
In \cref{fig_Flongt}, we present the same dynamics presented in \cref{fig_dynamics} of the main text, now up to $t=30$, nearly ten times longer than previously shown.
At this extended duration, we still observe persistent revivals and oscillations in both return fidelity and occupancy.
Moreover, in addition to the rapid oscillations with a period $T<1$ discussed in the main text, we observe a beating with frequency $T\sim 1$ attributed to the non-uniform distribution of scars in energy spacing.
Finally, we still observe a slow 
growth in entanglement at such extended times, consistent with dynamics constrained to the many-body scars subset of the spectrum.

\section{Numerical methods}\label{AppC}
\subsection{Matrix product state simulation}\label{AppMPS}
To solve the dynamics described by the model \cref{eq_Heff} we employ a numerical simulation based on Matrix product states (MPS).

Consider a quantum many-body system defined on a one-dimensional lattice with $N$ sites.
A generic pure quantum state $\ket{\psi}$ of this system lives in the tensor product $\mathcal{H} = \mathcal{H}_1 \otimes \mathcal{H}_2 \otimes \dots \otimes \mathcal{H}_N$ of $N$ local Hilbert spaces $\mathcal{H}_{j}$, each one being of finite dimension $d$.
This state $\ket{\psi}$ can be expanded as $\ket{\psi}=\sum_{i_1,..i_N}\psi_{i_1,i_2,..i_N}\ket{i_1,i_2,..i_N}$, where $\{\ket{i}_j\}$ is the canonical basis of the site $j$.
Providing an exact description of this state entails knowing all the $d^N$ complex coefficients $\psi_{i_1,i_2,..i_N}$.
From a computational point of view, this becomes infeasible for large values of $N$ due to the exponential increase of the memory and resources you would need.

The main idea of the MPS representation consists in reshaping the generic quantum state $\ket{\psi}$ of the one-dimensional system into the form \cite{1992CMaPh.144..443F, A_Klumper_1991, A_Klumper_1993}:
\begin{equation}
    \ket{\psi}=\sum_{i_1,..i_N} \mathrm{Tr} \left [ A^{(i_1)}_{1} A^{(i_2)}_{2}...A^{(i_N)}_{N} \right ] \ket{i_1,i_2,..i_N},
\end{equation}
where, for each specific set of physical indices $\{i_1,i_2,..i_N\}$, the trace of the product of the $A^{(i_j)}_{j}$ matrices gives back the state coefficient $\psi_{ i_1, i_2,.. i_N}$. Each matrix $A^{(i_j)}_j$ has dimension $D_{j-1}{\times} D_{j}$ and, when working with open-boundary conditions, finite-edge constraints are used, i.e. $D_{1}=1$ and $D_{N}=1$.
In principle, every state of a one-dimensional quantum many-body system can be written exactly in the MPS form, with bond dimensions $D_j$ growing exponentially with the lattice sites $N$ \cite{vidal2003efficient}. However, when dealing with low-energy states of local Hamiltonians, and their real-time dynamics for small-to-intermediate time, an approximate but efficient representation at fixed maximum bond dimensions is usually very accurate in capturing the essential features of the systems \cite{ORUS2014117, RevModPhys.82.277}. In this case, the number of parameters in the MPS representation grows only polynomially with the system's size, drastically reducing the overall computational complexity.

Bond dimension $D_j$ are crucial parameters since they allow interpolating between separable states (all $D_j = 1$) and strongly entangled states ($D_j \gg 1$). Thus, they estimate the quantum correlations and entanglement present in the MPS representation.  This implies that dynamical problems characterized by a slow entanglement entropy growth, as in the many-body scars dynamics, can be efficiently and accurately described by MPS ansatz with small or intermediate bond dimensions~\cite{verstraete2008matrix,schollwock2011density}.

To compute the time evolution of the system, we made use of state-of-the-art algorithms for MPS dynamical simulations: time-evolving block decimation (TEBD) algorithm \cite{PhysRevLett.93.040502} or the time-dependent variational principle (TDVP) \cite{PhysRevLett.107.070601} preferring the first to capture scars dynamics, with a precise time step resolution, and the second to capture the more computationally demanding ergodic dynamics. For a detailed description of the methods, please see \cite{PAECKEL2019167998}.
With both methods during the time evolution, we impose a fixed tolerance, ${\rm tol}$, for the truncation in the employed Schmidt decomposition.
The maximum bond dimension employed in the simulation to ensure such precision during the time evolution is upper bounded by a maximum bond dimension $D_{\rm max}$.
These quantities define the accuracy of the simulation and are reported in the Figure's caption.

\subsection{Exact diagonalization}
To obtain the full spectrum $\{\ket{\Phi_s}\}$ of the model given in \cref{eq_H} and extract the overlap and the entanglement behaviors displayed in \cref{fig_tower}, we use an Exact Diagonalization (ED) code \cite{Cataldi2024} that exploits all the main symmetries of the model.

For instance, on each link, we can resolve the $\mathbb{Z}_{2}$ symmetry obtained from the gauge link decomposition into a pair of rishons, and constrain the same electric Casimir on both sides.
Throughout all the simulations, we also select the zero baryon number density sector, $\hat{N_b}=\frac{1}{2}\sum_n (\hat{M}_n-1)=0$, corresponding to the half-filling case.

Ultimately, in the case of periodic boundary conditions (PBC), we also exploit translational invariance \cite{Sandvik2010} and restrict simulations within momentum sector $k=0$, which states such as the bare vacuum in \cref{eq_vacuum_state} and polarized vacuum \cref{eq_pol_vacuum_state} belong to.
Notice that, due to the staggered mass term in \cref{eq_H}, the model is invariant under translations of two sites.

\bibliography{biblio}

\begin{thebibliography}{85}%
\makeatletter
\providecommand \@ifxundefined [1]{%
 \@ifx{#1\undefined}
}%
\providecommand \@ifnum [1]{%
 \ifnum #1\expandafter \@firstoftwo
 \else \expandafter \@secondoftwo
 \fi
}%
\providecommand \@ifx [1]{%
 \ifx #1\expandafter \@firstoftwo
 \else \expandafter \@secondoftwo
 \fi
}%
\providecommand \natexlab [1]{#1}%
\providecommand \enquote  [1]{``#1''}%
\providecommand \bibnamefont  [1]{#1}%
\providecommand \bibfnamefont [1]{#1}%
\providecommand \citenamefont [1]{#1}%
\providecommand \href@noop [0]{\@secondoftwo}%
\providecommand \href [0]{\begingroup \@sanitize@url \@href}%
\providecommand \@href[1]{\@@startlink{#1}\@@href}%
\providecommand \@@href[1]{\endgroup#1\@@endlink}%
\providecommand \@sanitize@url [0]{\catcode `\\12\catcode `\$12\catcode
  `\&12\catcode `\#12\catcode `\^12\catcode `\_12\catcode `\%12\relax}%
\providecommand \@@startlink[1]{}%
\providecommand \@@endlink[0]{}%
\providecommand \url  [0]{\begingroup\@sanitize@url \@url }%
\providecommand \@url [1]{\endgroup\@href {#1}{\urlprefix }}%
\providecommand \urlprefix  [0]{URL }%
\providecommand \Eprint [0]{\href }%
\providecommand \doibase [0]{https://doi.org/}%
\providecommand \selectlanguage [0]{\@gobble}%
\providecommand \bibinfo  [0]{\@secondoftwo}%
\providecommand \bibfield  [0]{\@secondoftwo}%
\providecommand \translation [1]{[#1]}%
\providecommand \BibitemOpen [0]{}%
\providecommand \bibitemStop [0]{}%
\providecommand \bibitemNoStop [0]{.\EOS\space}%
\providecommand \EOS [0]{\spacefactor3000\relax}%
\providecommand \BibitemShut  [1]{\csname bibitem#1\endcsname}%
\let\auto@bib@innerbib\@empty
\bibitem [{\citenamefont {Reimann}(2008)}]{Reimann2008}%
  \BibitemOpen
  \bibfield  {author} {\bibinfo {author} {\bibfnamefont {P.}~\bibnamefont
  {Reimann}},\ }\href {https://doi.org/10.1103/PhysRevLett.101.190403}
  {\bibfield  {journal} {\bibinfo  {journal} {Phys. Rev. Lett.}\ }\textbf
  {\bibinfo {volume} {101}},\ \bibinfo {pages} {190403} (\bibinfo {year}
  {2008})}\BibitemShut {NoStop}%
\bibitem [{\citenamefont {Reichental}\ \emph {et~al.}(2018)\citenamefont
  {Reichental}, \citenamefont {Klempner}, \citenamefont {Kafri},\ and\
  \citenamefont {Podolsky}}]{Reichental2018}%
  \BibitemOpen
  \bibfield  {author} {\bibinfo {author} {\bibfnamefont {I.}~\bibnamefont
  {Reichental}}, \bibinfo {author} {\bibfnamefont {A.}~\bibnamefont
  {Klempner}}, \bibinfo {author} {\bibfnamefont {Y.}~\bibnamefont {Kafri}},\
  and\ \bibinfo {author} {\bibfnamefont {D.}~\bibnamefont {Podolsky}},\ }\href
  {https://doi.org/10.1103/PhysRevB.97.134301} {\bibfield  {journal} {\bibinfo
  {journal} {Phys. Rev. B}\ }\textbf {\bibinfo {volume} {97}},\ \bibinfo
  {pages} {134301} (\bibinfo {year} {2018})}\BibitemShut {NoStop}%
\bibitem [{\citenamefont {Deutsch}(1991)}]{Deutsch1991}%
  \BibitemOpen
  \bibfield  {author} {\bibinfo {author} {\bibfnamefont {J.~M.}\ \bibnamefont
  {Deutsch}},\ }\href {https://doi.org/10.1103/PhysRevA.43.2046} {\bibfield
  {journal} {\bibinfo  {journal} {Phys. Rev. A}\ }\textbf {\bibinfo {volume}
  {43}},\ \bibinfo {pages} {2046} (\bibinfo {year} {1991})}\BibitemShut
  {NoStop}%
\bibitem [{\citenamefont {Srednicki}(1994)}]{Srednicki1994}%
  \BibitemOpen
  \bibfield  {author} {\bibinfo {author} {\bibfnamefont {M.}~\bibnamefont
  {Srednicki}},\ }\href {https://doi.org/10.1103/PhysRevE.50.888} {\bibfield
  {journal} {\bibinfo  {journal} {Phys. Rev. E}\ }\textbf {\bibinfo {volume}
  {50}},\ \bibinfo {pages} {888} (\bibinfo {year} {1994})}\BibitemShut
  {NoStop}%
\bibitem [{\citenamefont {D'Alessio}\ \emph {et~al.}(2016)\citenamefont
  {D'Alessio}, \citenamefont {Kafri}, \citenamefont {Polkovnikov},\ and\
  \citenamefont {Rigol}}]{Rigol_review}%
  \BibitemOpen
  \bibfield  {author} {\bibinfo {author} {\bibfnamefont {L.}~\bibnamefont
  {D'Alessio}}, \bibinfo {author} {\bibfnamefont {Y.}~\bibnamefont {Kafri}},
  \bibinfo {author} {\bibfnamefont {A.}~\bibnamefont {Polkovnikov}},\ and\
  \bibinfo {author} {\bibfnamefont {M.}~\bibnamefont {Rigol}},\ }\href
  {https://doi.org/10.1080/00018732.2016.1198134} {\bibfield  {journal}
  {\bibinfo  {journal} {Advances in Physics}\ }\textbf {\bibinfo {volume}
  {65}},\ \bibinfo {pages} {239} (\bibinfo {year} {2016})}\BibitemShut
  {NoStop}%
\bibitem [{\citenamefont {Deutsch}(2018)}]{Deutsch_review}%
  \BibitemOpen
  \bibfield  {author} {\bibinfo {author} {\bibfnamefont {J.~M.}\ \bibnamefont
  {Deutsch}},\ }\href {https://doi.org/10.1088/1361-6633/aac9f1} {\bibfield
  {journal} {\bibinfo  {journal} {Reports on Progress in Physics}\ }\textbf
  {\bibinfo {volume} {81}},\ \bibinfo {pages} {082001} (\bibinfo {year}
  {2018})}\BibitemShut {NoStop}%
\bibitem [{\citenamefont {Rigol}\ \emph {et~al.}(2008)\citenamefont {Rigol},
  \citenamefont {Dunjko},\ and\ \citenamefont {Olshanii}}]{Rigol_2008}%
  \BibitemOpen
  \bibfield  {author} {\bibinfo {author} {\bibfnamefont {M.}~\bibnamefont
  {Rigol}}, \bibinfo {author} {\bibfnamefont {V.}~\bibnamefont {Dunjko}},\ and\
  \bibinfo {author} {\bibfnamefont {M.}~\bibnamefont {Olshanii}},\ }\href
  {https://doi.org/10.1038/nature06838} {\bibfield  {journal} {\bibinfo
  {journal} {Nature}\ }\textbf {\bibinfo {volume} {452}},\ \bibinfo {pages}
  {854} (\bibinfo {year} {2008})}\BibitemShut {NoStop}%
\bibitem [{\citenamefont {Turner}\ \emph {et~al.}(2018)\citenamefont {Turner},
  \citenamefont {Michailidis}, \citenamefont {Abanin}, \citenamefont {Serbyn},\
  and\ \citenamefont {Papi{\'c}}}]{Turner2018}%
  \BibitemOpen
  \bibfield  {author} {\bibinfo {author} {\bibfnamefont {C.~J.}\ \bibnamefont
  {Turner}}, \bibinfo {author} {\bibfnamefont {A.~A.}\ \bibnamefont
  {Michailidis}}, \bibinfo {author} {\bibfnamefont {D.~A.}\ \bibnamefont
  {Abanin}}, \bibinfo {author} {\bibfnamefont {M.}~\bibnamefont {Serbyn}},\
  and\ \bibinfo {author} {\bibfnamefont {Z.}~\bibnamefont {Papi{\'c}}},\ }\href
  {https://doi.org/10.1038/s41567-018-0137-5} {\bibfield  {journal} {\bibinfo
  {journal} {Nature Physics}\ }\textbf {\bibinfo {volume} {14}},\ \bibinfo
  {pages} {745} (\bibinfo {year} {2018})}\BibitemShut {NoStop}%
\bibitem [{\citenamefont {Moudgalya}\ \emph
  {et~al.}(2018{\natexlab{a}})\citenamefont {Moudgalya}, \citenamefont
  {Regnault},\ and\ \citenamefont {Bernevig}}]{BernevigEnt}%
  \BibitemOpen
  \bibfield  {author} {\bibinfo {author} {\bibfnamefont {S.}~\bibnamefont
  {Moudgalya}}, \bibinfo {author} {\bibfnamefont {N.}~\bibnamefont
  {Regnault}},\ and\ \bibinfo {author} {\bibfnamefont {B.~A.}\ \bibnamefont
  {Bernevig}},\ }\href {https://doi.org/10.1103/PhysRevB.98.235156} {\bibfield
  {journal} {\bibinfo  {journal} {Phys. Rev. B}\ }\textbf {\bibinfo {volume}
  {98}},\ \bibinfo {pages} {235156} (\bibinfo {year}
  {2018}{\natexlab{a}})}\BibitemShut {NoStop}%
\bibitem [{\citenamefont {Schecter}\ and\ \citenamefont
  {Iadecola}(2019)}]{Schecter2019}%
  \BibitemOpen
  \bibfield  {author} {\bibinfo {author} {\bibfnamefont {M.}~\bibnamefont
  {Schecter}}\ and\ \bibinfo {author} {\bibfnamefont {T.}~\bibnamefont
  {Iadecola}},\ }\href {https://doi.org/10.1103/PhysRevLett.123.147201}
  {\bibfield  {journal} {\bibinfo  {journal} {Phys. Rev. Lett.}\ }\textbf
  {\bibinfo {volume} {123}},\ \bibinfo {pages} {147201} (\bibinfo {year}
  {2019})}\BibitemShut {NoStop}%
\bibitem [{\citenamefont {Lin}\ and\ \citenamefont
  {Motrunich}(2019)}]{lin2018exact}%
  \BibitemOpen
  \bibfield  {author} {\bibinfo {author} {\bibfnamefont {C.-J.}\ \bibnamefont
  {Lin}}\ and\ \bibinfo {author} {\bibfnamefont {O.~I.}\ \bibnamefont
  {Motrunich}},\ }\href {https://doi.org/10.1103/PhysRevLett.122.173401}
  {\bibfield  {journal} {\bibinfo  {journal} {Phys. Rev. Lett.}\ }\textbf
  {\bibinfo {volume} {122}},\ \bibinfo {pages} {173401} (\bibinfo {year}
  {2019})}\BibitemShut {NoStop}%
\bibitem [{\citenamefont {Sun}\ \emph {et~al.}(2023)\citenamefont {Sun},
  \citenamefont {Popov}, \citenamefont {Klebanov},\ and\ \citenamefont
  {Pakrouski}}]{PhysRevResearch.5.043208}%
  \BibitemOpen
  \bibfield  {author} {\bibinfo {author} {\bibfnamefont {Z.}~\bibnamefont
  {Sun}}, \bibinfo {author} {\bibfnamefont {F.~K.}\ \bibnamefont {Popov}},
  \bibinfo {author} {\bibfnamefont {I.~R.}\ \bibnamefont {Klebanov}},\ and\
  \bibinfo {author} {\bibfnamefont {K.}~\bibnamefont {Pakrouski}},\ }\href
  {https://doi.org/10.1103/PhysRevResearch.5.043208} {\bibfield  {journal}
  {\bibinfo  {journal} {Phys. Rev. Res.}\ }\textbf {\bibinfo {volume} {5}},\
  \bibinfo {pages} {043208} (\bibinfo {year} {2023})}\BibitemShut {NoStop}%
\bibitem [{\citenamefont {Moudgalya}\ \emph
  {et~al.}(2018{\natexlab{b}})\citenamefont {Moudgalya}, \citenamefont
  {Regnault},\ and\ \citenamefont {Bernevig}}]{PhysRevB.98.235156}%
  \BibitemOpen
  \bibfield  {author} {\bibinfo {author} {\bibfnamefont {S.}~\bibnamefont
  {Moudgalya}}, \bibinfo {author} {\bibfnamefont {N.}~\bibnamefont
  {Regnault}},\ and\ \bibinfo {author} {\bibfnamefont {B.~A.}\ \bibnamefont
  {Bernevig}},\ }\href {https://doi.org/10.1103/PhysRevB.98.235156} {\bibfield
  {journal} {\bibinfo  {journal} {Phys. Rev. B}\ }\textbf {\bibinfo {volume}
  {98}},\ \bibinfo {pages} {235156} (\bibinfo {year}
  {2018}{\natexlab{b}})}\BibitemShut {NoStop}%
\bibitem [{\citenamefont {Bernien}\ \emph {et~al.}(2017)\citenamefont
  {Bernien}, \citenamefont {Schwartz}, \citenamefont {Keesling}, \citenamefont
  {Levine}, \citenamefont {Omran}, \citenamefont {Pichler}, \citenamefont
  {Choi}, \citenamefont {Zibrov}, \citenamefont {Endres}, \citenamefont
  {Greiner}, \citenamefont {Vuleti{\'c}},\ and\ \citenamefont
  {Lukin}}]{Bernien2017}%
  \BibitemOpen
  \bibfield  {author} {\bibinfo {author} {\bibfnamefont {H.}~\bibnamefont
  {Bernien}}, \bibinfo {author} {\bibfnamefont {S.}~\bibnamefont {Schwartz}},
  \bibinfo {author} {\bibfnamefont {A.}~\bibnamefont {Keesling}}, \bibinfo
  {author} {\bibfnamefont {H.}~\bibnamefont {Levine}}, \bibinfo {author}
  {\bibfnamefont {A.}~\bibnamefont {Omran}}, \bibinfo {author} {\bibfnamefont
  {H.}~\bibnamefont {Pichler}}, \bibinfo {author} {\bibfnamefont
  {S.}~\bibnamefont {Choi}}, \bibinfo {author} {\bibfnamefont {A.~S.}\
  \bibnamefont {Zibrov}}, \bibinfo {author} {\bibfnamefont {M.}~\bibnamefont
  {Endres}}, \bibinfo {author} {\bibfnamefont {M.}~\bibnamefont {Greiner}},
  \bibinfo {author} {\bibfnamefont {V.}~\bibnamefont {Vuleti{\'c}}},\ and\
  \bibinfo {author} {\bibfnamefont {M.~D.}\ \bibnamefont {Lukin}},\ }\href
  {https://doi.org/10.1038/nature24622} {\bibfield  {journal} {\bibinfo
  {journal} {Nature}\ }\textbf {\bibinfo {volume} {551}},\ \bibinfo {pages}
  {579} (\bibinfo {year} {2017})}\BibitemShut {NoStop}%
\bibitem [{\citenamefont {Moudgalya}\ \emph
  {et~al.}(2018{\natexlab{c}})\citenamefont {Moudgalya}, \citenamefont
  {Rachel}, \citenamefont {Bernevig},\ and\ \citenamefont
  {Regnault}}]{Moudgalya2018}%
  \BibitemOpen
  \bibfield  {author} {\bibinfo {author} {\bibfnamefont {S.}~\bibnamefont
  {Moudgalya}}, \bibinfo {author} {\bibfnamefont {S.}~\bibnamefont {Rachel}},
  \bibinfo {author} {\bibfnamefont {B.~A.}\ \bibnamefont {Bernevig}},\ and\
  \bibinfo {author} {\bibfnamefont {N.}~\bibnamefont {Regnault}},\ }\href
  {https://doi.org/10.1103/PhysRevB.98.235155} {\bibfield  {journal} {\bibinfo
  {journal} {Phys. Rev. B}\ }\textbf {\bibinfo {volume} {98}},\ \bibinfo
  {pages} {235155} (\bibinfo {year} {2018}{\natexlab{c}})}\BibitemShut
  {NoStop}%
\bibitem [{\citenamefont {Zhao}\ \emph {et~al.}(2020)\citenamefont {Zhao},
  \citenamefont {Vovrosh}, \citenamefont {Mintert},\ and\ \citenamefont
  {Knolle}}]{Zhao2020}%
  \BibitemOpen
  \bibfield  {author} {\bibinfo {author} {\bibfnamefont {H.}~\bibnamefont
  {Zhao}}, \bibinfo {author} {\bibfnamefont {J.}~\bibnamefont {Vovrosh}},
  \bibinfo {author} {\bibfnamefont {F.}~\bibnamefont {Mintert}},\ and\ \bibinfo
  {author} {\bibfnamefont {J.}~\bibnamefont {Knolle}},\ }\href
  {https://doi.org/10.1103/PhysRevLett.124.160604} {\bibfield  {journal}
  {\bibinfo  {journal} {Phys. Rev. Lett.}\ }\textbf {\bibinfo {volume} {124}},\
  \bibinfo {pages} {160604} (\bibinfo {year} {2020})}\BibitemShut {NoStop}%
\bibitem [{\citenamefont {Jepsen}\ \emph {et~al.}(2022)\citenamefont {Jepsen},
  \citenamefont {Lee}, \citenamefont {Lin}, \citenamefont {Dimitrova},
  \citenamefont {Margalit}, \citenamefont {Ho},\ and\ \citenamefont
  {Ketterle}}]{Jepsen2021}%
  \BibitemOpen
  \bibfield  {author} {\bibinfo {author} {\bibfnamefont {P.~N.}\ \bibnamefont
  {Jepsen}}, \bibinfo {author} {\bibfnamefont {Y.~K.~E.}\ \bibnamefont {Lee}},
  \bibinfo {author} {\bibfnamefont {H.}~\bibnamefont {Lin}}, \bibinfo {author}
  {\bibfnamefont {I.}~\bibnamefont {Dimitrova}}, \bibinfo {author}
  {\bibfnamefont {Y.}~\bibnamefont {Margalit}}, \bibinfo {author}
  {\bibfnamefont {W.~W.}\ \bibnamefont {Ho}},\ and\ \bibinfo {author}
  {\bibfnamefont {W.}~\bibnamefont {Ketterle}},\ }\href
  {https://doi.org/10.1038/s41567-022-01651-7} {\bibfield  {journal} {\bibinfo
  {journal} {Nature Physics}\ }\textbf {\bibinfo {volume} {18}},\ \bibinfo
  {pages} {899} (\bibinfo {year} {2022})}\BibitemShut {NoStop}%
\bibitem [{\citenamefont {Serbyn}\ \emph {et~al.}(2021)\citenamefont {Serbyn},
  \citenamefont {Abanin},\ and\ \citenamefont {Papi{\'c}}}]{Serbyn2020}%
  \BibitemOpen
  \bibfield  {author} {\bibinfo {author} {\bibfnamefont {M.}~\bibnamefont
  {Serbyn}}, \bibinfo {author} {\bibfnamefont {D.~A.}\ \bibnamefont {Abanin}},\
  and\ \bibinfo {author} {\bibfnamefont {Z.}~\bibnamefont {Papi{\'c}}},\ }\href
  {https://doi.org/10.1038/s41567-021-01230-2} {\bibfield  {journal} {\bibinfo
  {journal} {Nature Physics}\ }\textbf {\bibinfo {volume} {17}},\ \bibinfo
  {pages} {675} (\bibinfo {year} {2021})}\BibitemShut {NoStop}%
\bibitem [{\citenamefont {Moudgalya}\ \emph {et~al.}(2022)\citenamefont
  {Moudgalya}, \citenamefont {Bernevig},\ and\ \citenamefont
  {Regnault}}]{Moudgalya_review}%
  \BibitemOpen
  \bibfield  {author} {\bibinfo {author} {\bibfnamefont {S.}~\bibnamefont
  {Moudgalya}}, \bibinfo {author} {\bibfnamefont {B.~A.}\ \bibnamefont
  {Bernevig}},\ and\ \bibinfo {author} {\bibfnamefont {N.}~\bibnamefont
  {Regnault}},\ }\href {https://doi.org/10.1088/1361-6633/ac73a0} {\bibfield
  {journal} {\bibinfo  {journal} {Reports on Progress in Physics}\ }\textbf
  {\bibinfo {volume} {85}},\ \bibinfo {pages} {086501} (\bibinfo {year}
  {2022})}\BibitemShut {NoStop}%
\bibitem [{\citenamefont {Chandran}\ \emph {et~al.}(2023)\citenamefont
  {Chandran}, \citenamefont {Iadecola}, \citenamefont {Khemani},\ and\
  \citenamefont {Moessner}}]{Chandran_review}%
  \BibitemOpen
  \bibfield  {author} {\bibinfo {author} {\bibfnamefont {A.}~\bibnamefont
  {Chandran}}, \bibinfo {author} {\bibfnamefont {T.}~\bibnamefont {Iadecola}},
  \bibinfo {author} {\bibfnamefont {V.}~\bibnamefont {Khemani}},\ and\ \bibinfo
  {author} {\bibfnamefont {R.}~\bibnamefont {Moessner}},\ }\href
  {https://doi.org/10.1146/annurev-conmatphys-031620-101617} {\bibfield
  {journal} {\bibinfo  {journal} {Annual Review of Condensed Matter Physics}\
  }\textbf {\bibinfo {volume} {14}},\ \bibinfo {pages} {443} (\bibinfo {year}
  {2023})}\BibitemShut {NoStop}%
\bibitem [{\citenamefont {Bluvstein}\ \emph {et~al.}(2021)\citenamefont
  {Bluvstein}, \citenamefont {Omran}, \citenamefont {Levine}, \citenamefont
  {Keesling}, \citenamefont {Semeghini}, \citenamefont {Ebadi}, \citenamefont
  {Wang}, \citenamefont {Michailidis}, \citenamefont {Maskara}, \citenamefont
  {Ho}, \citenamefont {Choi}, \citenamefont {Serbyn}, \citenamefont {Greiner},
  \citenamefont {Vuleti{\'{c}}},\ and\ \citenamefont {Lukin}}]{Bluvstein2021}%
  \BibitemOpen
  \bibfield  {author} {\bibinfo {author} {\bibfnamefont {D.}~\bibnamefont
  {Bluvstein}}, \bibinfo {author} {\bibfnamefont {A.}~\bibnamefont {Omran}},
  \bibinfo {author} {\bibfnamefont {H.}~\bibnamefont {Levine}}, \bibinfo
  {author} {\bibfnamefont {A.}~\bibnamefont {Keesling}}, \bibinfo {author}
  {\bibfnamefont {G.}~\bibnamefont {Semeghini}}, \bibinfo {author}
  {\bibfnamefont {S.}~\bibnamefont {Ebadi}}, \bibinfo {author} {\bibfnamefont
  {T.~T.}\ \bibnamefont {Wang}}, \bibinfo {author} {\bibfnamefont {A.~A.}\
  \bibnamefont {Michailidis}}, \bibinfo {author} {\bibfnamefont
  {N.}~\bibnamefont {Maskara}}, \bibinfo {author} {\bibfnamefont {W.~W.}\
  \bibnamefont {Ho}}, \bibinfo {author} {\bibfnamefont {S.}~\bibnamefont
  {Choi}}, \bibinfo {author} {\bibfnamefont {M.}~\bibnamefont {Serbyn}},
  \bibinfo {author} {\bibfnamefont {M.}~\bibnamefont {Greiner}}, \bibinfo
  {author} {\bibfnamefont {V.}~\bibnamefont {Vuleti{\'{c}}}},\ and\ \bibinfo
  {author} {\bibfnamefont {M.~D.}\ \bibnamefont {Lukin}},\ }\href
  {https://doi.org/10.1126/science.abg2530} {\bibfield  {journal} {\bibinfo
  {journal} {Science}\ }\textbf {\bibinfo {volume} {371}},\ \bibinfo {pages}
  {1355} (\bibinfo {year} {2021})}\BibitemShut {NoStop}%
\bibitem [{\citenamefont {Bluvstein}\ \emph {et~al.}(2022)\citenamefont
  {Bluvstein}, \citenamefont {Levine}, \citenamefont {Semeghini}, \citenamefont
  {Wang}, \citenamefont {Ebadi}, \citenamefont {Kalinowski}, \citenamefont
  {Keesling}, \citenamefont {Maskara}, \citenamefont {Pichler}, \citenamefont
  {Greiner}, \citenamefont {Vuleti{\'c}},\ and\ \citenamefont
  {Lukin}}]{Bluvstein2022quantum}%
  \BibitemOpen
  \bibfield  {author} {\bibinfo {author} {\bibfnamefont {D.}~\bibnamefont
  {Bluvstein}}, \bibinfo {author} {\bibfnamefont {H.}~\bibnamefont {Levine}},
  \bibinfo {author} {\bibfnamefont {G.}~\bibnamefont {Semeghini}}, \bibinfo
  {author} {\bibfnamefont {T.~T.}\ \bibnamefont {Wang}}, \bibinfo {author}
  {\bibfnamefont {S.}~\bibnamefont {Ebadi}}, \bibinfo {author} {\bibfnamefont
  {M.}~\bibnamefont {Kalinowski}}, \bibinfo {author} {\bibfnamefont
  {A.}~\bibnamefont {Keesling}}, \bibinfo {author} {\bibfnamefont
  {N.}~\bibnamefont {Maskara}}, \bibinfo {author} {\bibfnamefont
  {H.}~\bibnamefont {Pichler}}, \bibinfo {author} {\bibfnamefont
  {M.}~\bibnamefont {Greiner}}, \bibinfo {author} {\bibfnamefont
  {V.}~\bibnamefont {Vuleti{\'c}}},\ and\ \bibinfo {author} {\bibfnamefont
  {M.~D.}\ \bibnamefont {Lukin}},\ }\href
  {https://doi.org/10.1038/s41586-022-04592-6} {\bibfield  {journal} {\bibinfo
  {journal} {Nature}\ }\textbf {\bibinfo {volume} {604}},\ \bibinfo {pages}
  {451} (\bibinfo {year} {2022})}\BibitemShut {NoStop}%
\bibitem [{\citenamefont {Su}\ \emph {et~al.}(2023)\citenamefont {Su},
  \citenamefont {Sun}, \citenamefont {Hudomal}, \citenamefont {Desaules},
  \citenamefont {Zhou}, \citenamefont {Yang}, \citenamefont {Halimeh},
  \citenamefont {Yuan}, \citenamefont {Papi\ifmmode~\acute{c}\else
  \'{c}\fi{}},\ and\ \citenamefont {Pan}}]{Su2022}%
  \BibitemOpen
  \bibfield  {author} {\bibinfo {author} {\bibfnamefont {G.-X.}\ \bibnamefont
  {Su}}, \bibinfo {author} {\bibfnamefont {H.}~\bibnamefont {Sun}}, \bibinfo
  {author} {\bibfnamefont {A.}~\bibnamefont {Hudomal}}, \bibinfo {author}
  {\bibfnamefont {J.-Y.}\ \bibnamefont {Desaules}}, \bibinfo {author}
  {\bibfnamefont {Z.-Y.}\ \bibnamefont {Zhou}}, \bibinfo {author}
  {\bibfnamefont {B.}~\bibnamefont {Yang}}, \bibinfo {author} {\bibfnamefont
  {J.~C.}\ \bibnamefont {Halimeh}}, \bibinfo {author} {\bibfnamefont {Z.-S.}\
  \bibnamefont {Yuan}}, \bibinfo {author} {\bibfnamefont {Z.}~\bibnamefont
  {Papi\ifmmode~\acute{c}\else \'{c}\fi{}}},\ and\ \bibinfo {author}
  {\bibfnamefont {J.-W.}\ \bibnamefont {Pan}},\ }\href
  {https://doi.org/10.1103/PhysRevResearch.5.023010} {\bibfield  {journal}
  {\bibinfo  {journal} {Phys. Rev. Res.}\ }\textbf {\bibinfo {volume} {5}},\
  \bibinfo {pages} {023010} (\bibinfo {year} {2023})}\BibitemShut {NoStop}%
\bibitem [{\citenamefont {Zhang}\ \emph {et~al.}(2023)\citenamefont {Zhang},
  \citenamefont {Dong}, \citenamefont {Gao}, \citenamefont {Zhao},
  \citenamefont {Hao}, \citenamefont {Desaules}, \citenamefont {Guo},
  \citenamefont {Chen}, \citenamefont {Deng}, \citenamefont {Liu},
  \citenamefont {Ren}, \citenamefont {Yao}, \citenamefont {Zhang},
  \citenamefont {Xu}, \citenamefont {Wang}, \citenamefont {Jin}, \citenamefont
  {Zhu}, \citenamefont {Zhang}, \citenamefont {Li}, \citenamefont {Song},
  \citenamefont {Wang}, \citenamefont {Liu}, \citenamefont {Papi{\'c}},
  \citenamefont {Ying}, \citenamefont {Wang},\ and\ \citenamefont
  {Lai}}]{Zhang2023Many-body}%
  \BibitemOpen
  \bibfield  {author} {\bibinfo {author} {\bibfnamefont {P.}~\bibnamefont
  {Zhang}}, \bibinfo {author} {\bibfnamefont {H.}~\bibnamefont {Dong}},
  \bibinfo {author} {\bibfnamefont {Y.}~\bibnamefont {Gao}}, \bibinfo {author}
  {\bibfnamefont {L.}~\bibnamefont {Zhao}}, \bibinfo {author} {\bibfnamefont
  {J.}~\bibnamefont {Hao}}, \bibinfo {author} {\bibfnamefont {J.-Y.}\
  \bibnamefont {Desaules}}, \bibinfo {author} {\bibfnamefont {Q.}~\bibnamefont
  {Guo}}, \bibinfo {author} {\bibfnamefont {J.}~\bibnamefont {Chen}}, \bibinfo
  {author} {\bibfnamefont {J.}~\bibnamefont {Deng}}, \bibinfo {author}
  {\bibfnamefont {B.}~\bibnamefont {Liu}}, \bibinfo {author} {\bibfnamefont
  {W.}~\bibnamefont {Ren}}, \bibinfo {author} {\bibfnamefont {Y.}~\bibnamefont
  {Yao}}, \bibinfo {author} {\bibfnamefont {X.}~\bibnamefont {Zhang}}, \bibinfo
  {author} {\bibfnamefont {S.}~\bibnamefont {Xu}}, \bibinfo {author}
  {\bibfnamefont {K.}~\bibnamefont {Wang}}, \bibinfo {author} {\bibfnamefont
  {F.}~\bibnamefont {Jin}}, \bibinfo {author} {\bibfnamefont {X.}~\bibnamefont
  {Zhu}}, \bibinfo {author} {\bibfnamefont {B.}~\bibnamefont {Zhang}}, \bibinfo
  {author} {\bibfnamefont {H.}~\bibnamefont {Li}}, \bibinfo {author}
  {\bibfnamefont {C.}~\bibnamefont {Song}}, \bibinfo {author} {\bibfnamefont
  {Z.}~\bibnamefont {Wang}}, \bibinfo {author} {\bibfnamefont {F.}~\bibnamefont
  {Liu}}, \bibinfo {author} {\bibfnamefont {Z.}~\bibnamefont {Papi{\'c}}},
  \bibinfo {author} {\bibfnamefont {L.}~\bibnamefont {Ying}}, \bibinfo {author}
  {\bibfnamefont {H.}~\bibnamefont {Wang}},\ and\ \bibinfo {author}
  {\bibfnamefont {Y.-C.}\ \bibnamefont {Lai}},\ }\href
  {https://doi.org/10.1038/s41567-022-01784-9} {\bibfield  {journal} {\bibinfo
  {journal} {Nature Physics}\ }\textbf {\bibinfo {volume} {19}},\ \bibinfo
  {pages} {120} (\bibinfo {year} {2023})}\BibitemShut {NoStop}%
\bibitem [{\citenamefont {Dong}\ \emph {et~al.}(2023)\citenamefont {Dong},
  \citenamefont {Desaules}, \citenamefont {Gao}, \citenamefont {Wang},
  \citenamefont {Guo}, \citenamefont {Chen}, \citenamefont {Zou}, \citenamefont
  {Jin}, \citenamefont {Zhu}, \citenamefont {Zhang}, \citenamefont {Li},
  \citenamefont {Wang}, \citenamefont {Guo}, \citenamefont {Zhang},
  \citenamefont {Ying},\ and\ \citenamefont {Papi{\'{c}}}}]{Dong2023Disorder}%
  \BibitemOpen
  \bibfield  {author} {\bibinfo {author} {\bibfnamefont {H.}~\bibnamefont
  {Dong}}, \bibinfo {author} {\bibfnamefont {J.-Y.}\ \bibnamefont {Desaules}},
  \bibinfo {author} {\bibfnamefont {Y.}~\bibnamefont {Gao}}, \bibinfo {author}
  {\bibfnamefont {N.}~\bibnamefont {Wang}}, \bibinfo {author} {\bibfnamefont
  {Z.}~\bibnamefont {Guo}}, \bibinfo {author} {\bibfnamefont {J.}~\bibnamefont
  {Chen}}, \bibinfo {author} {\bibfnamefont {Y.}~\bibnamefont {Zou}}, \bibinfo
  {author} {\bibfnamefont {F.}~\bibnamefont {Jin}}, \bibinfo {author}
  {\bibfnamefont {X.}~\bibnamefont {Zhu}}, \bibinfo {author} {\bibfnamefont
  {P.}~\bibnamefont {Zhang}}, \bibinfo {author} {\bibfnamefont
  {H.}~\bibnamefont {Li}}, \bibinfo {author} {\bibfnamefont {Z.}~\bibnamefont
  {Wang}}, \bibinfo {author} {\bibfnamefont {Q.}~\bibnamefont {Guo}}, \bibinfo
  {author} {\bibfnamefont {J.}~\bibnamefont {Zhang}}, \bibinfo {author}
  {\bibfnamefont {L.}~\bibnamefont {Ying}},\ and\ \bibinfo {author}
  {\bibfnamefont {Z.}~\bibnamefont {Papi{\'{c}}}},\ }\href
  {https://doi.org/10.1126/sciadv.adj3822} {\bibfield  {journal} {\bibinfo
  {journal} {Science Advances}\ }\textbf {\bibinfo {volume} {9}},\ \bibinfo
  {pages} {eadj3822} (\bibinfo {year} {2023})}\BibitemShut {NoStop}%
\bibitem [{\citenamefont {Surace}\ \emph {et~al.}(2020)\citenamefont {Surace},
  \citenamefont {Mazza}, \citenamefont {Giudici}, \citenamefont {Lerose},
  \citenamefont {Gambassi},\ and\ \citenamefont {Dalmonte}}]{Surace2020}%
  \BibitemOpen
  \bibfield  {author} {\bibinfo {author} {\bibfnamefont {F.~M.}\ \bibnamefont
  {Surace}}, \bibinfo {author} {\bibfnamefont {P.~P.}\ \bibnamefont {Mazza}},
  \bibinfo {author} {\bibfnamefont {G.}~\bibnamefont {Giudici}}, \bibinfo
  {author} {\bibfnamefont {A.}~\bibnamefont {Lerose}}, \bibinfo {author}
  {\bibfnamefont {A.}~\bibnamefont {Gambassi}},\ and\ \bibinfo {author}
  {\bibfnamefont {M.}~\bibnamefont {Dalmonte}},\ }\href
  {https://doi.org/10.1103/PhysRevX.10.021041} {\bibfield  {journal} {\bibinfo
  {journal} {Phys. Rev. X}\ }\textbf {\bibinfo {volume} {10}},\ \bibinfo
  {pages} {021041} (\bibinfo {year} {2020})}\BibitemShut {NoStop}%
\bibitem [{\citenamefont {Halimeh}\ \emph
  {et~al.}(2023{\natexlab{a}})\citenamefont {Halimeh}, \citenamefont
  {Barbiero}, \citenamefont {Hauke}, \citenamefont {Grusdt},\ and\
  \citenamefont {Bohrdt}}]{Halimeh2022robust}%
  \BibitemOpen
  \bibfield  {author} {\bibinfo {author} {\bibfnamefont {J.~C.}\ \bibnamefont
  {Halimeh}}, \bibinfo {author} {\bibfnamefont {L.}~\bibnamefont {Barbiero}},
  \bibinfo {author} {\bibfnamefont {P.}~\bibnamefont {Hauke}}, \bibinfo
  {author} {\bibfnamefont {F.}~\bibnamefont {Grusdt}},\ and\ \bibinfo {author}
  {\bibfnamefont {A.}~\bibnamefont {Bohrdt}},\ }\href
  {https://doi.org/10.22331/q-2023-05-15-1004} {\bibfield  {journal} {\bibinfo
  {journal} {{Quantum}}\ }\textbf {\bibinfo {volume} {7}},\ \bibinfo {pages}
  {1004} (\bibinfo {year} {2023}{\natexlab{a}})}\BibitemShut {NoStop}%
\bibitem [{\citenamefont {Desaules}\ \emph
  {et~al.}(2023{\natexlab{a}})\citenamefont {Desaules}, \citenamefont
  {Banerjee}, \citenamefont {Hudomal}, \citenamefont
  {Papi\ifmmode~\acute{c}\else \'{c}\fi{}}, \citenamefont {Sen},\ and\
  \citenamefont {Halimeh}}]{Desaules2022weak}%
  \BibitemOpen
  \bibfield  {author} {\bibinfo {author} {\bibfnamefont {J.-Y.}\ \bibnamefont
  {Desaules}}, \bibinfo {author} {\bibfnamefont {D.}~\bibnamefont {Banerjee}},
  \bibinfo {author} {\bibfnamefont {A.}~\bibnamefont {Hudomal}}, \bibinfo
  {author} {\bibfnamefont {Z.}~\bibnamefont {Papi\ifmmode~\acute{c}\else
  \'{c}\fi{}}}, \bibinfo {author} {\bibfnamefont {A.}~\bibnamefont {Sen}},\
  and\ \bibinfo {author} {\bibfnamefont {J.~C.}\ \bibnamefont {Halimeh}},\
  }\href {https://doi.org/10.1103/PhysRevB.107.L201105} {\bibfield  {journal}
  {\bibinfo  {journal} {Phys. Rev. B}\ }\textbf {\bibinfo {volume} {107}},\
  \bibinfo {pages} {L201105} (\bibinfo {year}
  {2023}{\natexlab{a}})}\BibitemShut {NoStop}%
\bibitem [{\citenamefont {Desaules}\ \emph
  {et~al.}(2023{\natexlab{b}})\citenamefont {Desaules}, \citenamefont
  {Hudomal}, \citenamefont {Banerjee}, \citenamefont {Sen}, \citenamefont
  {Papi\ifmmode~\acute{c}\else \'{c}\fi{}},\ and\ \citenamefont
  {Halimeh}}]{Desaules2022prominent}%
  \BibitemOpen
  \bibfield  {author} {\bibinfo {author} {\bibfnamefont {J.-Y.}\ \bibnamefont
  {Desaules}}, \bibinfo {author} {\bibfnamefont {A.}~\bibnamefont {Hudomal}},
  \bibinfo {author} {\bibfnamefont {D.}~\bibnamefont {Banerjee}}, \bibinfo
  {author} {\bibfnamefont {A.}~\bibnamefont {Sen}}, \bibinfo {author}
  {\bibfnamefont {Z.}~\bibnamefont {Papi\ifmmode~\acute{c}\else \'{c}\fi{}}},\
  and\ \bibinfo {author} {\bibfnamefont {J.~C.}\ \bibnamefont {Halimeh}},\
  }\href {https://doi.org/10.1103/PhysRevB.107.205112} {\bibfield  {journal}
  {\bibinfo  {journal} {Phys. Rev. B}\ }\textbf {\bibinfo {volume} {107}},\
  \bibinfo {pages} {205112} (\bibinfo {year} {2023}{\natexlab{b}})}\BibitemShut
  {NoStop}%
\bibitem [{\citenamefont {Iadecola}\ and\ \citenamefont
  {Schecter}(2020)}]{Iadecola2020quantum}%
  \BibitemOpen
  \bibfield  {author} {\bibinfo {author} {\bibfnamefont {T.}~\bibnamefont
  {Iadecola}}\ and\ \bibinfo {author} {\bibfnamefont {M.}~\bibnamefont
  {Schecter}},\ }\href {https://doi.org/10.1103/PhysRevB.101.024306} {\bibfield
   {journal} {\bibinfo  {journal} {Phys. Rev. B}\ }\textbf {\bibinfo {volume}
  {101}},\ \bibinfo {pages} {024306} (\bibinfo {year} {2020})}\BibitemShut
  {NoStop}%
\bibitem [{\citenamefont {Aramthottil}\ \emph {et~al.}(2022)\citenamefont
  {Aramthottil}, \citenamefont {Bhattacharya}, \citenamefont
  {Gonz\'alez-Cuadra}, \citenamefont {Lewenstein}, \citenamefont {Barbiero},\
  and\ \citenamefont {Zakrzewski}}]{aramthottil2022scar}%
  \BibitemOpen
  \bibfield  {author} {\bibinfo {author} {\bibfnamefont {A.~S.}\ \bibnamefont
  {Aramthottil}}, \bibinfo {author} {\bibfnamefont {U.}~\bibnamefont
  {Bhattacharya}}, \bibinfo {author} {\bibfnamefont {D.}~\bibnamefont
  {Gonz\'alez-Cuadra}}, \bibinfo {author} {\bibfnamefont {M.}~\bibnamefont
  {Lewenstein}}, \bibinfo {author} {\bibfnamefont {L.}~\bibnamefont
  {Barbiero}},\ and\ \bibinfo {author} {\bibfnamefont {J.}~\bibnamefont
  {Zakrzewski}},\ }\href {https://doi.org/10.1103/PhysRevB.106.L041101}
  {\bibfield  {journal} {\bibinfo  {journal} {Phys. Rev. B}\ }\textbf {\bibinfo
  {volume} {106}},\ \bibinfo {pages} {L041101} (\bibinfo {year}
  {2022})}\BibitemShut {NoStop}%
\bibitem [{\citenamefont {Desaules}\ \emph {et~al.}(2024)\citenamefont
  {Desaules}, \citenamefont {Iadecola},\ and\ \citenamefont
  {Halimeh}}]{desaules2024massassisted}%
  \BibitemOpen
  \bibfield  {author} {\bibinfo {author} {\bibfnamefont {J.-Y.}\ \bibnamefont
  {Desaules}}, \bibinfo {author} {\bibfnamefont {T.}~\bibnamefont {Iadecola}},\
  and\ \bibinfo {author} {\bibfnamefont {J.~C.}\ \bibnamefont {Halimeh}},\
  }\bibfield  {journal} {\bibinfo  {journal} {arXiv eprints}\ }\href
  {https://doi.org/10.48550/arxiv.2404.11645} {10.48550/arxiv.2404.11645}
  (\bibinfo {year} {2024}),\ \Eprint {https://arxiv.org/abs/2404.11645}
  {arXiv:2404.11645 [cond-mat.quant-gas]} \BibitemShut {NoStop}%
\bibitem [{\citenamefont {Banerjee}\ and\ \citenamefont
  {Sen}(2021)}]{Banerjee2021}%
  \BibitemOpen
  \bibfield  {author} {\bibinfo {author} {\bibfnamefont {D.}~\bibnamefont
  {Banerjee}}\ and\ \bibinfo {author} {\bibfnamefont {A.}~\bibnamefont {Sen}},\
  }\href {https://doi.org/10.1103/PhysRevLett.126.220601} {\bibfield  {journal}
  {\bibinfo  {journal} {Phys. Rev. Lett.}\ }\textbf {\bibinfo {volume} {126}},\
  \bibinfo {pages} {220601} (\bibinfo {year} {2021})}\BibitemShut {NoStop}%
\bibitem [{\citenamefont {Biswas}\ \emph {et~al.}(2022)\citenamefont {Biswas},
  \citenamefont {Banerjee},\ and\ \citenamefont {Sen}}]{biswas2022scars}%
  \BibitemOpen
  \bibfield  {author} {\bibinfo {author} {\bibfnamefont {S.}~\bibnamefont
  {Biswas}}, \bibinfo {author} {\bibfnamefont {D.}~\bibnamefont {Banerjee}},\
  and\ \bibinfo {author} {\bibfnamefont {A.}~\bibnamefont {Sen}},\ }\href
  {https://doi.org/10.21468/SciPostPhys.12.5.148} {\bibfield  {journal}
  {\bibinfo  {journal} {SciPost Phys.}\ }\textbf {\bibinfo {volume} {12}},\
  \bibinfo {pages} {148} (\bibinfo {year} {2022})}\BibitemShut {NoStop}%
\bibitem [{\citenamefont {Ebner}\ \emph {et~al.}(2024)\citenamefont {Ebner},
  \citenamefont {Sch\"afer}, \citenamefont {Seidl}, \citenamefont {M\"uller},\
  and\ \citenamefont {Yao}}]{ebner2024entanglement}%
  \BibitemOpen
  \bibfield  {author} {\bibinfo {author} {\bibfnamefont {L.}~\bibnamefont
  {Ebner}}, \bibinfo {author} {\bibfnamefont {A.}~\bibnamefont {Sch\"afer}},
  \bibinfo {author} {\bibfnamefont {C.}~\bibnamefont {Seidl}}, \bibinfo
  {author} {\bibfnamefont {B.}~\bibnamefont {M\"uller}},\ and\ \bibinfo
  {author} {\bibfnamefont {X.}~\bibnamefont {Yao}},\ }\href
  {https://doi.org/10.1103/PhysRevD.110.014505} {\bibfield  {journal} {\bibinfo
   {journal} {Phys. Rev. D}\ }\textbf {\bibinfo {volume} {110}},\ \bibinfo
  {pages} {014505} (\bibinfo {year} {2024})}\BibitemShut {NoStop}%
\bibitem [{\citenamefont {Sau}\ \emph {et~al.}(2024)\citenamefont {Sau},
  \citenamefont {Stornati}, \citenamefont {Banerjee},\ and\ \citenamefont
  {Sen}}]{Sau2024}%
  \BibitemOpen
  \bibfield  {author} {\bibinfo {author} {\bibfnamefont {I.}~\bibnamefont
  {Sau}}, \bibinfo {author} {\bibfnamefont {P.}~\bibnamefont {Stornati}},
  \bibinfo {author} {\bibfnamefont {D.}~\bibnamefont {Banerjee}},\ and\
  \bibinfo {author} {\bibfnamefont {A.}~\bibnamefont {Sen}},\ }\href
  {https://doi.org/10.1103/PhysRevD.109.034519} {\bibfield  {journal} {\bibinfo
   {journal} {Phys. Rev. D}\ }\textbf {\bibinfo {volume} {109}},\ \bibinfo
  {pages} {034519} (\bibinfo {year} {2024})}\BibitemShut {NoStop}%
\bibitem [{\citenamefont {Osborne}\ \emph {et~al.}(2024)\citenamefont
  {Osborne}, \citenamefont {McCulloch},\ and\ \citenamefont
  {Halimeh}}]{osborne2024quantum}%
  \BibitemOpen
  \bibfield  {author} {\bibinfo {author} {\bibfnamefont {J.}~\bibnamefont
  {Osborne}}, \bibinfo {author} {\bibfnamefont {I.~P.}\ \bibnamefont
  {McCulloch}},\ and\ \bibinfo {author} {\bibfnamefont {J.~C.}\ \bibnamefont
  {Halimeh}},\ }\bibfield  {journal} {\bibinfo  {journal} {arXiv eprints}\
  }\href {https://doi.org/10.48550/arxiv.2403.08858}
  {10.48550/arxiv.2403.08858} (\bibinfo {year} {2024}),\ \Eprint
  {https://arxiv.org/abs/2403.08858} {arXiv:2403.08858 [cond-mat.quant-gas]}
  \BibitemShut {NoStop}%
\bibitem [{\citenamefont {Budde}\ \emph
  {et~al.}(2024{\natexlab{a}})\citenamefont {Budde}, \citenamefont
  {Marinkovi{\'{c}}},\ and\ \citenamefont {Barros}}]{budde2024quantum}%
  \BibitemOpen
  \bibfield  {author} {\bibinfo {author} {\bibfnamefont {T.}~\bibnamefont
  {Budde}}, \bibinfo {author} {\bibfnamefont {M.~K.}\ \bibnamefont
  {Marinkovi{\'{c}}}},\ and\ \bibinfo {author} {\bibfnamefont {J.~C.~P.}\
  \bibnamefont {Barros}},\ }\bibfield  {journal} {\bibinfo  {journal} {arXiv
  eprints}\ }\href {https://doi.org/10.48550/arxiv.2403.08892}
  {10.48550/arxiv.2403.08892} (\bibinfo {year} {2024}{\natexlab{a}}),\ \Eprint
  {https://arxiv.org/abs/2403.08892} {arXiv:2403.08892 [hep-lat]} \BibitemShut
  {NoStop}%
\bibitem [{\citenamefont {Dalmonte}\ and\ \citenamefont
  {Montangero}(2016)}]{Dalmonte_review}%
  \BibitemOpen
  \bibfield  {author} {\bibinfo {author} {\bibfnamefont {M.}~\bibnamefont
  {Dalmonte}}\ and\ \bibinfo {author} {\bibfnamefont {S.}~\bibnamefont
  {Montangero}},\ }\href {https://doi.org/10.1080/00107514.2016.1151199}
  {\bibfield  {journal} {\bibinfo  {journal} {Contemporary Physics}\ }\textbf
  {\bibinfo {volume} {57}},\ \bibinfo {pages} {388} (\bibinfo {year}
  {2016})}\BibitemShut {NoStop}%
\bibitem [{\citenamefont {Ba{\~n}uls}\ \emph {et~al.}(2020)\citenamefont
  {Ba{\~n}uls}, \citenamefont {Blatt}, \citenamefont {Catani}, \citenamefont
  {Celi}, \citenamefont {Cirac}, \citenamefont {Dalmonte}, \citenamefont
  {Fallani}, \citenamefont {Jansen}, \citenamefont {Lewenstein}, \citenamefont
  {Montangero}, \citenamefont {Muschik}, \citenamefont {Reznik}, \citenamefont
  {Rico}, \citenamefont {Tagliacozzo}, \citenamefont {Van~Acoleyen},
  \citenamefont {Verstraete}, \citenamefont {Wiese}, \citenamefont {Wingate},
  \citenamefont {Zakrzewski},\ and\ \citenamefont {Zoller}}]{Pasquans_review}%
  \BibitemOpen
  \bibfield  {author} {\bibinfo {author} {\bibfnamefont {M.~C.}\ \bibnamefont
  {Ba{\~n}uls}}, \bibinfo {author} {\bibfnamefont {R.}~\bibnamefont {Blatt}},
  \bibinfo {author} {\bibfnamefont {J.}~\bibnamefont {Catani}}, \bibinfo
  {author} {\bibfnamefont {A.}~\bibnamefont {Celi}}, \bibinfo {author}
  {\bibfnamefont {J.~I.}\ \bibnamefont {Cirac}}, \bibinfo {author}
  {\bibfnamefont {M.}~\bibnamefont {Dalmonte}}, \bibinfo {author}
  {\bibfnamefont {L.}~\bibnamefont {Fallani}}, \bibinfo {author} {\bibfnamefont
  {K.}~\bibnamefont {Jansen}}, \bibinfo {author} {\bibfnamefont
  {M.}~\bibnamefont {Lewenstein}}, \bibinfo {author} {\bibfnamefont
  {S.}~\bibnamefont {Montangero}}, \bibinfo {author} {\bibfnamefont {C.~A.}\
  \bibnamefont {Muschik}}, \bibinfo {author} {\bibfnamefont {B.}~\bibnamefont
  {Reznik}}, \bibinfo {author} {\bibfnamefont {E.}~\bibnamefont {Rico}},
  \bibinfo {author} {\bibfnamefont {L.}~\bibnamefont {Tagliacozzo}}, \bibinfo
  {author} {\bibfnamefont {K.}~\bibnamefont {Van~Acoleyen}}, \bibinfo {author}
  {\bibfnamefont {F.}~\bibnamefont {Verstraete}}, \bibinfo {author}
  {\bibfnamefont {U.-J.}\ \bibnamefont {Wiese}}, \bibinfo {author}
  {\bibfnamefont {M.}~\bibnamefont {Wingate}}, \bibinfo {author} {\bibfnamefont
  {J.}~\bibnamefont {Zakrzewski}},\ and\ \bibinfo {author} {\bibfnamefont
  {P.}~\bibnamefont {Zoller}},\ }\href
  {https://doi.org/10.1140/epjd/e2020-100571-8} {\bibfield  {journal} {\bibinfo
   {journal} {The European Physical Journal D}\ }\textbf {\bibinfo {volume}
  {74}},\ \bibinfo {pages} {165} (\bibinfo {year} {2020})}\BibitemShut
  {NoStop}%
\bibitem [{\citenamefont {Zohar}\ \emph {et~al.}(2015)\citenamefont {Zohar},
  \citenamefont {Cirac},\ and\ \citenamefont {Reznik}}]{Zohar_review}%
  \BibitemOpen
  \bibfield  {author} {\bibinfo {author} {\bibfnamefont {E.}~\bibnamefont
  {Zohar}}, \bibinfo {author} {\bibfnamefont {J.~I.}\ \bibnamefont {Cirac}},\
  and\ \bibinfo {author} {\bibfnamefont {B.}~\bibnamefont {Reznik}},\ }\href
  {https://doi.org/10.1088/0034-4885/79/1/014401} {\bibfield  {journal}
  {\bibinfo  {journal} {Reports on Progress in Physics}\ }\textbf {\bibinfo
  {volume} {79}},\ \bibinfo {pages} {014401} (\bibinfo {year}
  {2015})}\BibitemShut {NoStop}%
\bibitem [{\citenamefont {Alexeev}\ \emph {et~al.}(2021)\citenamefont
  {Alexeev}, \citenamefont {Bacon}, \citenamefont {Brown}, \citenamefont
  {Calderbank}, \citenamefont {Carr}, \citenamefont {Chong}, \citenamefont
  {DeMarco}, \citenamefont {Englund}, \citenamefont {Farhi}, \citenamefont
  {Fefferman}, \citenamefont {Gorshkov}, \citenamefont {Houck}, \citenamefont
  {Kim}, \citenamefont {Kimmel}, \citenamefont {Lange}, \citenamefont {Lloyd},
  \citenamefont {Lukin}, \citenamefont {Maslov}, \citenamefont {Maunz},
  \citenamefont {Monroe}, \citenamefont {Preskill}, \citenamefont {Roetteler},
  \citenamefont {Savage},\ and\ \citenamefont {Thompson}}]{Alexeev_review}%
  \BibitemOpen
  \bibfield  {author} {\bibinfo {author} {\bibfnamefont {Y.}~\bibnamefont
  {Alexeev}}, \bibinfo {author} {\bibfnamefont {D.}~\bibnamefont {Bacon}},
  \bibinfo {author} {\bibfnamefont {K.~R.}\ \bibnamefont {Brown}}, \bibinfo
  {author} {\bibfnamefont {R.}~\bibnamefont {Calderbank}}, \bibinfo {author}
  {\bibfnamefont {L.~D.}\ \bibnamefont {Carr}}, \bibinfo {author}
  {\bibfnamefont {F.~T.}\ \bibnamefont {Chong}}, \bibinfo {author}
  {\bibfnamefont {B.}~\bibnamefont {DeMarco}}, \bibinfo {author} {\bibfnamefont
  {D.}~\bibnamefont {Englund}}, \bibinfo {author} {\bibfnamefont
  {E.}~\bibnamefont {Farhi}}, \bibinfo {author} {\bibfnamefont
  {B.}~\bibnamefont {Fefferman}}, \bibinfo {author} {\bibfnamefont {A.~V.}\
  \bibnamefont {Gorshkov}}, \bibinfo {author} {\bibfnamefont {A.}~\bibnamefont
  {Houck}}, \bibinfo {author} {\bibfnamefont {J.}~\bibnamefont {Kim}}, \bibinfo
  {author} {\bibfnamefont {S.}~\bibnamefont {Kimmel}}, \bibinfo {author}
  {\bibfnamefont {M.}~\bibnamefont {Lange}}, \bibinfo {author} {\bibfnamefont
  {S.}~\bibnamefont {Lloyd}}, \bibinfo {author} {\bibfnamefont {M.~D.}\
  \bibnamefont {Lukin}}, \bibinfo {author} {\bibfnamefont {D.}~\bibnamefont
  {Maslov}}, \bibinfo {author} {\bibfnamefont {P.}~\bibnamefont {Maunz}},
  \bibinfo {author} {\bibfnamefont {C.}~\bibnamefont {Monroe}}, \bibinfo
  {author} {\bibfnamefont {J.}~\bibnamefont {Preskill}}, \bibinfo {author}
  {\bibfnamefont {M.}~\bibnamefont {Roetteler}}, \bibinfo {author}
  {\bibfnamefont {M.~J.}\ \bibnamefont {Savage}},\ and\ \bibinfo {author}
  {\bibfnamefont {J.}~\bibnamefont {Thompson}},\ }\href
  {https://doi.org/10.1103/PRXQuantum.2.017001} {\bibfield  {journal} {\bibinfo
   {journal} {PRX Quantum}\ }\textbf {\bibinfo {volume} {2}},\ \bibinfo {pages}
  {017001} (\bibinfo {year} {2021})}\BibitemShut {NoStop}%
\bibitem [{\citenamefont {Aidelsburger}\ \emph {et~al.}(2022)\citenamefont
  {Aidelsburger}, \citenamefont {Barbiero}, \citenamefont {Bermudez},
  \citenamefont {Chanda}, \citenamefont {Dauphin}, \citenamefont
  {Gonz{\'{a}}lez-Cuadra}, \citenamefont {Grzybowski}, \citenamefont {Hands},
  \citenamefont {Jendrzejewski}, \citenamefont {J{\"{u}}nemann}, \citenamefont
  {Juzeli{\={u}}nas}, \citenamefont {Kasper}, \citenamefont {Piga},
  \citenamefont {Ran}, \citenamefont {Rizzi}, \citenamefont {Sierra},
  \citenamefont {Tagliacozzo}, \citenamefont {Tirrito}, \citenamefont {Zache},
  \citenamefont {Zakrzewski}, \citenamefont {Zohar},\ and\ \citenamefont
  {Lewenstein}}]{aidelsburger2021cold}%
  \BibitemOpen
  \bibfield  {author} {\bibinfo {author} {\bibfnamefont {M.}~\bibnamefont
  {Aidelsburger}}, \bibinfo {author} {\bibfnamefont {L.}~\bibnamefont
  {Barbiero}}, \bibinfo {author} {\bibfnamefont {A.}~\bibnamefont {Bermudez}},
  \bibinfo {author} {\bibfnamefont {T.}~\bibnamefont {Chanda}}, \bibinfo
  {author} {\bibfnamefont {A.}~\bibnamefont {Dauphin}}, \bibinfo {author}
  {\bibfnamefont {D.}~\bibnamefont {Gonz{\'{a}}lez-Cuadra}}, \bibinfo {author}
  {\bibfnamefont {P.~R.}\ \bibnamefont {Grzybowski}}, \bibinfo {author}
  {\bibfnamefont {S.}~\bibnamefont {Hands}}, \bibinfo {author} {\bibfnamefont
  {F.}~\bibnamefont {Jendrzejewski}}, \bibinfo {author} {\bibfnamefont
  {J.}~\bibnamefont {J{\"{u}}nemann}}, \bibinfo {author} {\bibfnamefont
  {G.}~\bibnamefont {Juzeli{\={u}}nas}}, \bibinfo {author} {\bibfnamefont
  {V.}~\bibnamefont {Kasper}}, \bibinfo {author} {\bibfnamefont
  {A.}~\bibnamefont {Piga}}, \bibinfo {author} {\bibfnamefont {S.-J.}\
  \bibnamefont {Ran}}, \bibinfo {author} {\bibfnamefont {M.}~\bibnamefont
  {Rizzi}}, \bibinfo {author} {\bibfnamefont {G.}~\bibnamefont {Sierra}},
  \bibinfo {author} {\bibfnamefont {L.}~\bibnamefont {Tagliacozzo}}, \bibinfo
  {author} {\bibfnamefont {E.}~\bibnamefont {Tirrito}}, \bibinfo {author}
  {\bibfnamefont {T.~V.}\ \bibnamefont {Zache}}, \bibinfo {author}
  {\bibfnamefont {J.}~\bibnamefont {Zakrzewski}}, \bibinfo {author}
  {\bibfnamefont {E.}~\bibnamefont {Zohar}},\ and\ \bibinfo {author}
  {\bibfnamefont {M.}~\bibnamefont {Lewenstein}},\ }\href
  {https://doi.org/10.1098/rsta.2021.0064} {\bibfield  {journal} {\bibinfo
  {journal} {Philosophical Transactions of the Royal Society A: Mathematical,
  Physical and Engineering Sciences}\ }\textbf {\bibinfo {volume} {380}},\
  \bibinfo {pages} {20210064} (\bibinfo {year} {2022})}\BibitemShut {NoStop}%
\bibitem [{\citenamefont {Zohar}(2022)}]{zohar2021quantum}%
  \BibitemOpen
  \bibfield  {author} {\bibinfo {author} {\bibfnamefont {E.}~\bibnamefont
  {Zohar}},\ }\href {https://doi.org/10.1098/rsta.2021.0069} {\bibfield
  {journal} {\bibinfo  {journal} {Philosophical Transactions of the Royal
  Society A: Mathematical, Physical and Engineering Sciences}\ }\textbf
  {\bibinfo {volume} {380}},\ \bibinfo {pages} {20210069} (\bibinfo {year}
  {2022})}\BibitemShut {NoStop}%
\bibitem [{\citenamefont {Klco}\ \emph {et~al.}(2022)\citenamefont {Klco},
  \citenamefont {Roggero},\ and\ \citenamefont {Savage}}]{klco2021standard}%
  \BibitemOpen
  \bibfield  {author} {\bibinfo {author} {\bibfnamefont {N.}~\bibnamefont
  {Klco}}, \bibinfo {author} {\bibfnamefont {A.}~\bibnamefont {Roggero}},\ and\
  \bibinfo {author} {\bibfnamefont {M.~J.}\ \bibnamefont {Savage}},\ }\href
  {https://doi.org/10.1088/1361-6633/ac58a4} {\bibfield  {journal} {\bibinfo
  {journal} {Reports on Progress in Physics}\ }\textbf {\bibinfo {volume}
  {85}},\ \bibinfo {pages} {064301} (\bibinfo {year} {2022})}\BibitemShut
  {NoStop}%
\bibitem [{\citenamefont {Bauer}\ \emph {et~al.}(2023)\citenamefont {Bauer},
  \citenamefont {Davoudi}, \citenamefont {Balantekin}, \citenamefont
  {Bhattacharya}, \citenamefont {Carena}, \citenamefont {de~Jong},
  \citenamefont {Draper}, \citenamefont {El-Khadra}, \citenamefont {Gemelke},
  \citenamefont {Hanada}, \citenamefont {Kharzeev}, \citenamefont {Lamm},
  \citenamefont {Li}, \citenamefont {Liu}, \citenamefont {Lukin}, \citenamefont
  {Meurice}, \citenamefont {Monroe}, \citenamefont {Nachman}, \citenamefont
  {Pagano}, \citenamefont {Preskill}, \citenamefont {Rinaldi}, \citenamefont
  {Roggero}, \citenamefont {Santiago}, \citenamefont {Savage}, \citenamefont
  {Siddiqi}, \citenamefont {Siopsis}, \citenamefont {Van~Zanten}, \citenamefont
  {Wiebe}, \citenamefont {Yamauchi}, \citenamefont {Yeter-Aydeniz},\ and\
  \citenamefont {Zorzetti}}]{Bauer_review}%
  \BibitemOpen
  \bibfield  {author} {\bibinfo {author} {\bibfnamefont {C.~W.}\ \bibnamefont
  {Bauer}}, \bibinfo {author} {\bibfnamefont {Z.}~\bibnamefont {Davoudi}},
  \bibinfo {author} {\bibfnamefont {A.~B.}\ \bibnamefont {Balantekin}},
  \bibinfo {author} {\bibfnamefont {T.}~\bibnamefont {Bhattacharya}}, \bibinfo
  {author} {\bibfnamefont {M.}~\bibnamefont {Carena}}, \bibinfo {author}
  {\bibfnamefont {W.~A.}\ \bibnamefont {de~Jong}}, \bibinfo {author}
  {\bibfnamefont {P.}~\bibnamefont {Draper}}, \bibinfo {author} {\bibfnamefont
  {A.}~\bibnamefont {El-Khadra}}, \bibinfo {author} {\bibfnamefont
  {N.}~\bibnamefont {Gemelke}}, \bibinfo {author} {\bibfnamefont
  {M.}~\bibnamefont {Hanada}}, \bibinfo {author} {\bibfnamefont
  {D.}~\bibnamefont {Kharzeev}}, \bibinfo {author} {\bibfnamefont
  {H.}~\bibnamefont {Lamm}}, \bibinfo {author} {\bibfnamefont {Y.-Y.}\
  \bibnamefont {Li}}, \bibinfo {author} {\bibfnamefont {J.}~\bibnamefont
  {Liu}}, \bibinfo {author} {\bibfnamefont {M.}~\bibnamefont {Lukin}}, \bibinfo
  {author} {\bibfnamefont {Y.}~\bibnamefont {Meurice}}, \bibinfo {author}
  {\bibfnamefont {C.}~\bibnamefont {Monroe}}, \bibinfo {author} {\bibfnamefont
  {B.}~\bibnamefont {Nachman}}, \bibinfo {author} {\bibfnamefont
  {G.}~\bibnamefont {Pagano}}, \bibinfo {author} {\bibfnamefont
  {J.}~\bibnamefont {Preskill}}, \bibinfo {author} {\bibfnamefont
  {E.}~\bibnamefont {Rinaldi}}, \bibinfo {author} {\bibfnamefont
  {A.}~\bibnamefont {Roggero}}, \bibinfo {author} {\bibfnamefont {D.~I.}\
  \bibnamefont {Santiago}}, \bibinfo {author} {\bibfnamefont {M.~J.}\
  \bibnamefont {Savage}}, \bibinfo {author} {\bibfnamefont {I.}~\bibnamefont
  {Siddiqi}}, \bibinfo {author} {\bibfnamefont {G.}~\bibnamefont {Siopsis}},
  \bibinfo {author} {\bibfnamefont {D.}~\bibnamefont {Van~Zanten}}, \bibinfo
  {author} {\bibfnamefont {N.}~\bibnamefont {Wiebe}}, \bibinfo {author}
  {\bibfnamefont {Y.}~\bibnamefont {Yamauchi}}, \bibinfo {author}
  {\bibfnamefont {K.}~\bibnamefont {Yeter-Aydeniz}},\ and\ \bibinfo {author}
  {\bibfnamefont {S.}~\bibnamefont {Zorzetti}},\ }\href
  {https://doi.org/10.1103/PRXQuantum.4.027001} {\bibfield  {journal} {\bibinfo
   {journal} {PRX Quantum}\ }\textbf {\bibinfo {volume} {4}},\ \bibinfo {pages}
  {027001} (\bibinfo {year} {2023})}\BibitemShut {NoStop}%
\bibitem [{\citenamefont {Di~Meglio}\ \emph {et~al.}(2024)\citenamefont
  {Di~Meglio}, \citenamefont {Jansen}, \citenamefont {Tavernelli},
  \citenamefont {Alexandrou}, \citenamefont {Arunachalam}, \citenamefont
  {Bauer}, \citenamefont {Borras}, \citenamefont {Carrazza}, \citenamefont
  {Crippa}, \citenamefont {Croft}, \citenamefont {de~Putter}, \citenamefont
  {Delgado}, \citenamefont {Dunjko}, \citenamefont {Egger}, \citenamefont
  {Fern\'andez-Combarro}, \citenamefont {Fuchs}, \citenamefont {Funcke},
  \citenamefont {Gonz\'alez-Cuadra}, \citenamefont {Grossi}, \citenamefont
  {Halimeh}, \citenamefont {Holmes}, \citenamefont {K\"uhn}, \citenamefont
  {Lacroix}, \citenamefont {Lewis}, \citenamefont {Lucchesi}, \citenamefont
  {Martinez}, \citenamefont {Meloni}, \citenamefont {Mezzacapo}, \citenamefont
  {Montangero}, \citenamefont {Nagano}, \citenamefont {Pascuzzi}, \citenamefont
  {Radescu}, \citenamefont {Ortega}, \citenamefont {Roggero}, \citenamefont
  {Schuhmacher}, \citenamefont {Seixas}, \citenamefont {Silvi}, \citenamefont
  {Spentzouris}, \citenamefont {Tacchino}, \citenamefont {Temme}, \citenamefont
  {Terashi}, \citenamefont {Tura}, \citenamefont {T\"uys\"uz}, \citenamefont
  {Vallecorsa}, \citenamefont {Wiese}, \citenamefont {Yoo},\ and\ \citenamefont
  {Zhang}}]{dimeglio2023quantum}%
  \BibitemOpen
  \bibfield  {author} {\bibinfo {author} {\bibfnamefont {A.}~\bibnamefont
  {Di~Meglio}}, \bibinfo {author} {\bibfnamefont {K.}~\bibnamefont {Jansen}},
  \bibinfo {author} {\bibfnamefont {I.}~\bibnamefont {Tavernelli}}, \bibinfo
  {author} {\bibfnamefont {C.}~\bibnamefont {Alexandrou}}, \bibinfo {author}
  {\bibfnamefont {S.}~\bibnamefont {Arunachalam}}, \bibinfo {author}
  {\bibfnamefont {C.~W.}\ \bibnamefont {Bauer}}, \bibinfo {author}
  {\bibfnamefont {K.}~\bibnamefont {Borras}}, \bibinfo {author} {\bibfnamefont
  {S.}~\bibnamefont {Carrazza}}, \bibinfo {author} {\bibfnamefont
  {A.}~\bibnamefont {Crippa}}, \bibinfo {author} {\bibfnamefont
  {V.}~\bibnamefont {Croft}}, \bibinfo {author} {\bibfnamefont
  {R.}~\bibnamefont {de~Putter}}, \bibinfo {author} {\bibfnamefont
  {A.}~\bibnamefont {Delgado}}, \bibinfo {author} {\bibfnamefont
  {V.}~\bibnamefont {Dunjko}}, \bibinfo {author} {\bibfnamefont {D.~J.}\
  \bibnamefont {Egger}}, \bibinfo {author} {\bibfnamefont {E.}~\bibnamefont
  {Fern\'andez-Combarro}}, \bibinfo {author} {\bibfnamefont {E.}~\bibnamefont
  {Fuchs}}, \bibinfo {author} {\bibfnamefont {L.}~\bibnamefont {Funcke}},
  \bibinfo {author} {\bibfnamefont {D.}~\bibnamefont {Gonz\'alez-Cuadra}},
  \bibinfo {author} {\bibfnamefont {M.}~\bibnamefont {Grossi}}, \bibinfo
  {author} {\bibfnamefont {J.~C.}\ \bibnamefont {Halimeh}}, \bibinfo {author}
  {\bibfnamefont {Z.}~\bibnamefont {Holmes}}, \bibinfo {author} {\bibfnamefont
  {S.}~\bibnamefont {K\"uhn}}, \bibinfo {author} {\bibfnamefont
  {D.}~\bibnamefont {Lacroix}}, \bibinfo {author} {\bibfnamefont
  {R.}~\bibnamefont {Lewis}}, \bibinfo {author} {\bibfnamefont
  {D.}~\bibnamefont {Lucchesi}}, \bibinfo {author} {\bibfnamefont {M.~L.}\
  \bibnamefont {Martinez}}, \bibinfo {author} {\bibfnamefont {F.}~\bibnamefont
  {Meloni}}, \bibinfo {author} {\bibfnamefont {A.}~\bibnamefont {Mezzacapo}},
  \bibinfo {author} {\bibfnamefont {S.}~\bibnamefont {Montangero}}, \bibinfo
  {author} {\bibfnamefont {L.}~\bibnamefont {Nagano}}, \bibinfo {author}
  {\bibfnamefont {V.~R.}\ \bibnamefont {Pascuzzi}}, \bibinfo {author}
  {\bibfnamefont {V.}~\bibnamefont {Radescu}}, \bibinfo {author} {\bibfnamefont
  {E.~R.}\ \bibnamefont {Ortega}}, \bibinfo {author} {\bibfnamefont
  {A.}~\bibnamefont {Roggero}}, \bibinfo {author} {\bibfnamefont
  {J.}~\bibnamefont {Schuhmacher}}, \bibinfo {author} {\bibfnamefont
  {J.}~\bibnamefont {Seixas}}, \bibinfo {author} {\bibfnamefont
  {P.}~\bibnamefont {Silvi}}, \bibinfo {author} {\bibfnamefont
  {P.}~\bibnamefont {Spentzouris}}, \bibinfo {author} {\bibfnamefont
  {F.}~\bibnamefont {Tacchino}}, \bibinfo {author} {\bibfnamefont
  {K.}~\bibnamefont {Temme}}, \bibinfo {author} {\bibfnamefont
  {K.}~\bibnamefont {Terashi}}, \bibinfo {author} {\bibfnamefont
  {J.}~\bibnamefont {Tura}}, \bibinfo {author} {\bibfnamefont {C.}~\bibnamefont
  {T\"uys\"uz}}, \bibinfo {author} {\bibfnamefont {S.}~\bibnamefont
  {Vallecorsa}}, \bibinfo {author} {\bibfnamefont {U.-J.}\ \bibnamefont
  {Wiese}}, \bibinfo {author} {\bibfnamefont {S.}~\bibnamefont {Yoo}},\ and\
  \bibinfo {author} {\bibfnamefont {J.}~\bibnamefont {Zhang}},\ }\href
  {https://doi.org/10.1103/PRXQuantum.5.037001} {\bibfield  {journal} {\bibinfo
   {journal} {PRX Quantum}\ }\textbf {\bibinfo {volume} {5}},\ \bibinfo {pages}
  {037001} (\bibinfo {year} {2024})}\BibitemShut {NoStop}%
\bibitem [{\citenamefont {Halimeh}\ \emph
  {et~al.}(2023{\natexlab{b}})\citenamefont {Halimeh}, \citenamefont
  {Aidelsburger}, \citenamefont {Grusdt}, \citenamefont {Hauke},\ and\
  \citenamefont {Yang}}]{halimeh2023coldatom}%
  \BibitemOpen
  \bibfield  {author} {\bibinfo {author} {\bibfnamefont {J.~C.}\ \bibnamefont
  {Halimeh}}, \bibinfo {author} {\bibfnamefont {M.}~\bibnamefont
  {Aidelsburger}}, \bibinfo {author} {\bibfnamefont {F.}~\bibnamefont
  {Grusdt}}, \bibinfo {author} {\bibfnamefont {P.}~\bibnamefont {Hauke}},\ and\
  \bibinfo {author} {\bibfnamefont {B.}~\bibnamefont {Yang}},\ }\bibfield
  {journal} {\bibinfo  {journal} {arXiv eprints}\ }\href
  {https://doi.org/10.48550/arxiv.2310.12201} {10.48550/arxiv.2310.12201}
  (\bibinfo {year} {2023}{\natexlab{b}}),\ \Eprint
  {https://arxiv.org/abs/2310.12201} {arXiv:2310.12201 [cond-mat.quant-gas]}
  \BibitemShut {NoStop}%
\bibitem [{\citenamefont {Cheng}\ and\ \citenamefont
  {Zhai}(2024)}]{cheng2024emergent}%
  \BibitemOpen
  \bibfield  {author} {\bibinfo {author} {\bibfnamefont {Y.}~\bibnamefont
  {Cheng}}\ and\ \bibinfo {author} {\bibfnamefont {H.}~\bibnamefont {Zhai}},\
  }\href
  {https://doi.org/https://www.nature.com/articles/s42254-024-00749-6#citeas}
  {\bibfield  {journal} {\bibinfo  {journal} {Nature Reviews Physics}\ }\textbf
  {\bibinfo {volume} {6}},\ \bibinfo {pages} {566} (\bibinfo {year}
  {2024})}\BibitemShut {NoStop}%
\bibitem [{\citenamefont {Cataldi}\ \emph {et~al.}(2024)\citenamefont
  {Cataldi}, \citenamefont {Magnifico}, \citenamefont {Silvi},\ and\
  \citenamefont {Montangero}}]{Cataldi2024SimulatingSUYangMills}%
  \BibitemOpen
  \bibfield  {author} {\bibinfo {author} {\bibfnamefont {G.}~\bibnamefont
  {Cataldi}}, \bibinfo {author} {\bibfnamefont {G.}~\bibnamefont {Magnifico}},
  \bibinfo {author} {\bibfnamefont {P.}~\bibnamefont {Silvi}},\ and\ \bibinfo
  {author} {\bibfnamefont {S.}~\bibnamefont {Montangero}},\ }\href
  {https://doi.org/10.1103/PhysRevResearch.6.033057} {\bibfield  {journal}
  {\bibinfo  {journal} {Physical Review Research}\ }\textbf {\bibinfo {volume}
  {6}},\ \bibinfo {pages} {033057} (\bibinfo {year} {2024})}\BibitemShut
  {NoStop}%
\bibitem [{\citenamefont {Rigobello}\ \emph {et~al.}(2023)\citenamefont
  {Rigobello}, \citenamefont {Magnifico}, \citenamefont {Silvi},\ and\
  \citenamefont {Montangero}}]{rigobello2023hadrons}%
  \BibitemOpen
  \bibfield  {author} {\bibinfo {author} {\bibfnamefont {M.}~\bibnamefont
  {Rigobello}}, \bibinfo {author} {\bibfnamefont {G.}~\bibnamefont
  {Magnifico}}, \bibinfo {author} {\bibfnamefont {P.}~\bibnamefont {Silvi}},\
  and\ \bibinfo {author} {\bibfnamefont {S.}~\bibnamefont {Montangero}},\
  }\bibfield  {journal} {\bibinfo  {journal} {arXiv eprints}\ }\href
  {https://doi.org/10.48550/arxiv.2308.04488} {10.48550/arxiv.2308.04488}
  (\bibinfo {year} {2023}),\ \Eprint {https://arxiv.org/abs/2308.04488}
  {arXiv:2308.04488 [hep-lat]} \BibitemShut {NoStop}%
\bibitem [{\citenamefont {Calaj{\'o}}\ \emph {et~al.}(2024)\citenamefont
  {Calaj{\'o}}, \citenamefont {Magnifico}, \citenamefont {Edmunds},
  \citenamefont {Ringbauer}, \citenamefont {Montangero},\ and\ \citenamefont
  {Silvi}}]{Calajo2024DigitalQuantumSimulation}%
  \BibitemOpen
  \bibfield  {author} {\bibinfo {author} {\bibfnamefont {G.}~\bibnamefont
  {Calaj{\'o}}}, \bibinfo {author} {\bibfnamefont {G.}~\bibnamefont
  {Magnifico}}, \bibinfo {author} {\bibfnamefont {C.}~\bibnamefont {Edmunds}},
  \bibinfo {author} {\bibfnamefont {M.}~\bibnamefont {Ringbauer}}, \bibinfo
  {author} {\bibfnamefont {S.}~\bibnamefont {Montangero}},\ and\ \bibinfo
  {author} {\bibfnamefont {P.}~\bibnamefont {Silvi}},\ }\href
  {https://doi.org/10.1103/PRXQuantum.5.040309} {\bibfield  {journal} {\bibinfo
   {journal} {PRX Quantum}\ }\textbf {\bibinfo {volume} {5}},\ \bibinfo {pages}
  {040309} (\bibinfo {year} {2024})}\BibitemShut {NoStop}%
\bibitem [{\citenamefont {Silvi}\ \emph {et~al.}(2019)\citenamefont {Silvi},
  \citenamefont {Sauer}, \citenamefont {Tschirsich},\ and\ \citenamefont
  {Montangero}}]{Silvi2019Tensor}%
  \BibitemOpen
  \bibfield  {author} {\bibinfo {author} {\bibfnamefont {P.}~\bibnamefont
  {Silvi}}, \bibinfo {author} {\bibfnamefont {Y.}~\bibnamefont {Sauer}},
  \bibinfo {author} {\bibfnamefont {F.}~\bibnamefont {Tschirsich}},\ and\
  \bibinfo {author} {\bibfnamefont {S.}~\bibnamefont {Montangero}},\ }\href
  {https://doi.org/10.1103/PhysRevD.100.074512} {\bibfield  {journal} {\bibinfo
   {journal} {Phys. Rev. D}\ }\textbf {\bibinfo {volume} {100}},\ \bibinfo
  {pages} {074512} (\bibinfo {year} {2019})}\BibitemShut {NoStop}%
\bibitem [{\citenamefont {Zohar}\ and\ \citenamefont
  {Cirac}(2019)}]{Zohar2019Removing}%
  \BibitemOpen
  \bibfield  {author} {\bibinfo {author} {\bibfnamefont {E.}~\bibnamefont
  {Zohar}}\ and\ \bibinfo {author} {\bibfnamefont {J.~I.}\ \bibnamefont
  {Cirac}},\ }\href {https://doi.org/10.1103/PhysRevD.99.114511} {\bibfield
  {journal} {\bibinfo  {journal} {Phys. Rev. D}\ }\textbf {\bibinfo {volume}
  {99}},\ \bibinfo {pages} {114511} (\bibinfo {year} {2019})}\BibitemShut
  {NoStop}%
\bibitem [{\citenamefont {Kogut}\ and\ \citenamefont
  {Susskind}(1975)}]{Kogut1975}%
  \BibitemOpen
  \bibfield  {author} {\bibinfo {author} {\bibfnamefont {J.}~\bibnamefont
  {Kogut}}\ and\ \bibinfo {author} {\bibfnamefont {L.}~\bibnamefont
  {Susskind}},\ }\href {https://doi.org/10.1103/PhysRevD.11.395} {\bibfield
  {journal} {\bibinfo  {journal} {Phys. Rev. D}\ }\textbf {\bibinfo {volume}
  {11}},\ \bibinfo {pages} {395} (\bibinfo {year} {1975})}\BibitemShut
  {NoStop}%
\bibitem [{\citenamefont {Susskind}(1977)}]{susskind1977lattice}%
  \BibitemOpen
  \bibfield  {author} {\bibinfo {author} {\bibfnamefont {L.}~\bibnamefont
  {Susskind}},\ }\href {https://doi.org/10.1103/PhysRevD.16.3031} {\bibfield
  {journal} {\bibinfo  {journal} {Phys. Rev. D}\ }\textbf {\bibinfo {volume}
  {16}},\ \bibinfo {pages} {3031} (\bibinfo {year} {1977})}\BibitemShut
  {NoStop}%
\bibitem [{\citenamefont {Zohar}\ and\ \citenamefont
  {Burrello}(2015)}]{Zohar2015}%
  \BibitemOpen
  \bibfield  {author} {\bibinfo {author} {\bibfnamefont {E.}~\bibnamefont
  {Zohar}}\ and\ \bibinfo {author} {\bibfnamefont {M.}~\bibnamefont
  {Burrello}},\ }\href {https://doi.org/10.1103/PhysRevD.91.054506} {\bibfield
  {journal} {\bibinfo  {journal} {Physical Review D}\ }\textbf {\bibinfo
  {volume} {91}},\ \bibinfo {pages} {054506} (\bibinfo {year}
  {2015})}\BibitemShut {NoStop}%
\bibitem [{\citenamefont {Banerjee}\ \emph {et~al.}(2012)\citenamefont
  {Banerjee}, \citenamefont {Dalmonte}, \citenamefont {M{\"{u}}ller},
  \citenamefont {Rico}, \citenamefont {Stebler}, \citenamefont {Wiese},\ and\
  \citenamefont {Zoller}}]{Banerjee2012}%
  \BibitemOpen
  \bibfield  {author} {\bibinfo {author} {\bibfnamefont {D.}~\bibnamefont
  {Banerjee}}, \bibinfo {author} {\bibfnamefont {M.}~\bibnamefont {Dalmonte}},
  \bibinfo {author} {\bibfnamefont {M.}~\bibnamefont {M{\"{u}}ller}}, \bibinfo
  {author} {\bibfnamefont {E.}~\bibnamefont {Rico}}, \bibinfo {author}
  {\bibfnamefont {P.}~\bibnamefont {Stebler}}, \bibinfo {author} {\bibfnamefont
  {U.-J.}\ \bibnamefont {Wiese}},\ and\ \bibinfo {author} {\bibfnamefont
  {P.}~\bibnamefont {Zoller}},\ }\bibfield  {journal} {\bibinfo  {journal}
  {Physical Review Letters}\ }\textbf {\bibinfo {volume} {109}},\ \href
  {https://doi.org/10.1103/physrevlett.109.175302}
  {10.1103/physrevlett.109.175302} (\bibinfo {year} {2012})\BibitemShut
  {NoStop}%
\bibitem [{\citenamefont {K\"uhn}\ \emph {et~al.}(2014)\citenamefont {K\"uhn},
  \citenamefont {Cirac},\ and\ \citenamefont {Ba\~nuls}}]{Kuehn2014}%
  \BibitemOpen
  \bibfield  {author} {\bibinfo {author} {\bibfnamefont {S.}~\bibnamefont
  {K\"uhn}}, \bibinfo {author} {\bibfnamefont {J.~I.}\ \bibnamefont {Cirac}},\
  and\ \bibinfo {author} {\bibfnamefont {M.-C.}\ \bibnamefont {Ba\~nuls}},\
  }\href {https://doi.org/10.1103/PhysRevA.90.042305} {\bibfield  {journal}
  {\bibinfo  {journal} {Phys. Rev. A}\ }\textbf {\bibinfo {volume} {90}},\
  \bibinfo {pages} {042305} (\bibinfo {year} {2014})}\BibitemShut {NoStop}%
\bibitem [{\citenamefont {Magnifico}\ \emph {et~al.}(2020)\citenamefont
  {Magnifico}, \citenamefont {Dalmonte}, \citenamefont {Facchi}, \citenamefont
  {Pascazio}, \citenamefont {Pepe},\ and\ \citenamefont
  {Ercolessi}}]{Magnifico2020realtimedynamics}%
  \BibitemOpen
  \bibfield  {author} {\bibinfo {author} {\bibfnamefont {G.}~\bibnamefont
  {Magnifico}}, \bibinfo {author} {\bibfnamefont {M.}~\bibnamefont {Dalmonte}},
  \bibinfo {author} {\bibfnamefont {P.}~\bibnamefont {Facchi}}, \bibinfo
  {author} {\bibfnamefont {S.}~\bibnamefont {Pascazio}}, \bibinfo {author}
  {\bibfnamefont {F.~V.}\ \bibnamefont {Pepe}},\ and\ \bibinfo {author}
  {\bibfnamefont {E.}~\bibnamefont {Ercolessi}},\ }\href
  {https://doi.org/10.22331/q-2020-06-15-281} {\bibfield  {journal} {\bibinfo
  {journal} {{Quantum}}\ }\textbf {\bibinfo {volume} {4}},\ \bibinfo {pages}
  {281} (\bibinfo {year} {2020})}\BibitemShut {NoStop}%
\bibitem [{\citenamefont {Daniel}\ \emph {et~al.}(2023)\citenamefont {Daniel},
  \citenamefont {Hallam}, \citenamefont {Desaules}, \citenamefont {Hudomal},
  \citenamefont {Su}, \citenamefont {Halimeh},\ and\ \citenamefont
  {Papi\ifmmode~\acute{c}\else \'{c}\fi{}}}]{Daniel2023}%
  \BibitemOpen
  \bibfield  {author} {\bibinfo {author} {\bibfnamefont {A.}~\bibnamefont
  {Daniel}}, \bibinfo {author} {\bibfnamefont {A.}~\bibnamefont {Hallam}},
  \bibinfo {author} {\bibfnamefont {J.-Y.}\ \bibnamefont {Desaules}}, \bibinfo
  {author} {\bibfnamefont {A.}~\bibnamefont {Hudomal}}, \bibinfo {author}
  {\bibfnamefont {G.-X.}\ \bibnamefont {Su}}, \bibinfo {author} {\bibfnamefont
  {J.~C.}\ \bibnamefont {Halimeh}},\ and\ \bibinfo {author} {\bibfnamefont
  {Z.}~\bibnamefont {Papi\ifmmode~\acute{c}\else \'{c}\fi{}}},\ }\href
  {https://doi.org/10.1103/PhysRevB.107.235108} {\bibfield  {journal} {\bibinfo
   {journal} {Phys. Rev. B}\ }\textbf {\bibinfo {volume} {107}},\ \bibinfo
  {pages} {235108} (\bibinfo {year} {2023})}\BibitemShut {NoStop}%
\bibitem [{\citenamefont {Hudomal}\ \emph {et~al.}(2022)\citenamefont
  {Hudomal}, \citenamefont {Desaules}, \citenamefont {Mukherjee}, \citenamefont
  {Su}, \citenamefont {Halimeh},\ and\ \citenamefont
  {Papi\ifmmode~\acute{c}\else \'{c}\fi{}}}]{Hudomal2022}%
  \BibitemOpen
  \bibfield  {author} {\bibinfo {author} {\bibfnamefont {A.}~\bibnamefont
  {Hudomal}}, \bibinfo {author} {\bibfnamefont {J.-Y.}\ \bibnamefont
  {Desaules}}, \bibinfo {author} {\bibfnamefont {B.}~\bibnamefont {Mukherjee}},
  \bibinfo {author} {\bibfnamefont {G.-X.}\ \bibnamefont {Su}}, \bibinfo
  {author} {\bibfnamefont {J.~C.}\ \bibnamefont {Halimeh}},\ and\ \bibinfo
  {author} {\bibfnamefont {Z.}~\bibnamefont {Papi\ifmmode~\acute{c}\else
  \'{c}\fi{}}},\ }\href {https://doi.org/10.1103/PhysRevB.106.104302}
  {\bibfield  {journal} {\bibinfo  {journal} {Phys. Rev. B}\ }\textbf {\bibinfo
  {volume} {106}},\ \bibinfo {pages} {104302} (\bibinfo {year}
  {2022})}\BibitemShut {NoStop}%
\bibitem [{\citenamefont {Moudgalya}\ and\ \citenamefont
  {Motrunich}(2024)}]{Moudgalya2023ExhaustiveCharacterizationQuantum}%
  \BibitemOpen
  \bibfield  {author} {\bibinfo {author} {\bibfnamefont {S.}~\bibnamefont
  {Moudgalya}}\ and\ \bibinfo {author} {\bibfnamefont {O.~I.}\ \bibnamefont
  {Motrunich}},\ }\href {https://doi.org/10.1103/PhysRevX.14.041069} {\bibfield
   {journal} {\bibinfo  {journal} {Phys. Rev. X}\ }\textbf {\bibinfo {volume}
  {14}},\ \bibinfo {pages} {041069} (\bibinfo {year} {2024})}\BibitemShut
  {NoStop}%
\bibitem [{\citenamefont {Budde}\ \emph
  {et~al.}(2024{\natexlab{b}})\citenamefont {Budde}, \citenamefont
  {Marinkovi{\'{c}}},\ and\ \citenamefont {Barros}}]{Budde2024}%
  \BibitemOpen
  \bibfield  {author} {\bibinfo {author} {\bibfnamefont {T.}~\bibnamefont
  {Budde}}, \bibinfo {author} {\bibfnamefont {M.~K.}\ \bibnamefont
  {Marinkovi{\'{c}}}},\ and\ \bibinfo {author} {\bibfnamefont {J.~C.~P.}\
  \bibnamefont {Barros}},\ }\bibfield  {journal} {\bibinfo  {journal} {arXiv
  eprints}\ }\href {https://doi.org/10.48550/arxiv.2403.08892}
  {10.48550/arxiv.2403.08892} (\bibinfo {year} {2024}{\natexlab{b}}),\ \Eprint
  {https://arxiv.org/abs/2403.08892} {arXiv:2403.08892 [hep-lat]} \BibitemShut
  {NoStop}%
\bibitem [{\citenamefont {Van~Damme}\ \emph {et~al.}(2022)\citenamefont
  {Van~Damme}, \citenamefont {Zache}, \citenamefont {Banerjee}, \citenamefont
  {Hauke},\ and\ \citenamefont {Halimeh}}]{VanDamme2022dynamical}%
  \BibitemOpen
  \bibfield  {author} {\bibinfo {author} {\bibfnamefont {M.}~\bibnamefont
  {Van~Damme}}, \bibinfo {author} {\bibfnamefont {T.~V.}\ \bibnamefont
  {Zache}}, \bibinfo {author} {\bibfnamefont {D.}~\bibnamefont {Banerjee}},
  \bibinfo {author} {\bibfnamefont {P.}~\bibnamefont {Hauke}},\ and\ \bibinfo
  {author} {\bibfnamefont {J.~C.}\ \bibnamefont {Halimeh}},\ }\href
  {https://doi.org/10.1103/PhysRevB.106.245110} {\bibfield  {journal} {\bibinfo
   {journal} {Phys. Rev. B}\ }\textbf {\bibinfo {volume} {106}},\ \bibinfo
  {pages} {245110} (\bibinfo {year} {2022})}\BibitemShut {NoStop}%
\bibitem [{\citenamefont {Van~Damme}\ \emph {et~al.}(2023)\citenamefont
  {Van~Damme}, \citenamefont {Desaules}, \citenamefont
  {Papi\ifmmode~\acute{c}\else \'{c}\fi{}},\ and\ \citenamefont
  {Halimeh}}]{VanDamme2023anatomy}%
  \BibitemOpen
  \bibfield  {author} {\bibinfo {author} {\bibfnamefont {M.}~\bibnamefont
  {Van~Damme}}, \bibinfo {author} {\bibfnamefont {J.-Y.}\ \bibnamefont
  {Desaules}}, \bibinfo {author} {\bibfnamefont {Z.}~\bibnamefont
  {Papi\ifmmode~\acute{c}\else \'{c}\fi{}}},\ and\ \bibinfo {author}
  {\bibfnamefont {J.~C.}\ \bibnamefont {Halimeh}},\ }\href
  {https://doi.org/10.1103/PhysRevResearch.5.033090} {\bibfield  {journal}
  {\bibinfo  {journal} {Phys. Rev. Res.}\ }\textbf {\bibinfo {volume} {5}},\
  \bibinfo {pages} {033090} (\bibinfo {year} {2023})}\BibitemShut {NoStop}%
\bibitem [{\citenamefont {Ringbauer}\ \emph {et~al.}(2022)\citenamefont
  {Ringbauer}, \citenamefont {Meth}, \citenamefont {Postler}, \citenamefont
  {Stricker}, \citenamefont {Blatt}, \citenamefont {Schindler},\ and\
  \citenamefont {Monz}}]{ringbauer2022universal}%
  \BibitemOpen
  \bibfield  {author} {\bibinfo {author} {\bibfnamefont {M.}~\bibnamefont
  {Ringbauer}}, \bibinfo {author} {\bibfnamefont {M.}~\bibnamefont {Meth}},
  \bibinfo {author} {\bibfnamefont {L.}~\bibnamefont {Postler}}, \bibinfo
  {author} {\bibfnamefont {R.}~\bibnamefont {Stricker}}, \bibinfo {author}
  {\bibfnamefont {R.}~\bibnamefont {Blatt}}, \bibinfo {author} {\bibfnamefont
  {P.}~\bibnamefont {Schindler}},\ and\ \bibinfo {author} {\bibfnamefont
  {T.}~\bibnamefont {Monz}},\ }\href
  {https://doi.org/https://doi.org/10.1038/s41567-022-01658-0} {\bibfield
  {journal} {\bibinfo  {journal} {Nature Physics}\ }\textbf {\bibinfo {volume}
  {18}},\ \bibinfo {pages} {1053} (\bibinfo {year} {2022})}\BibitemShut
  {NoStop}%
\bibitem [{\citenamefont {Fishman}\ \emph
  {et~al.}(2022{\natexlab{a}})\citenamefont {Fishman}, \citenamefont {White},\
  and\ \citenamefont {Stoudenmire}}]{ITensor}%
  \BibitemOpen
  \bibfield  {author} {\bibinfo {author} {\bibfnamefont {M.}~\bibnamefont
  {Fishman}}, \bibinfo {author} {\bibfnamefont {S.~R.}\ \bibnamefont {White}},\
  and\ \bibinfo {author} {\bibfnamefont {E.~M.}\ \bibnamefont {Stoudenmire}},\
  }\href {https://doi.org/10.21468/SciPostPhysCodeb.4} {\bibfield  {journal}
  {\bibinfo  {journal} {SciPost Phys. Codebases}\ ,\ \bibinfo {pages} {4}}
  (\bibinfo {year} {2022}{\natexlab{a}})}\BibitemShut {NoStop}%
\bibitem [{\citenamefont {Fishman}\ \emph
  {et~al.}(2022{\natexlab{b}})\citenamefont {Fishman}, \citenamefont {White},\
  and\ \citenamefont {Stoudenmire}}]{ITensor-r0.3}%
  \BibitemOpen
  \bibfield  {author} {\bibinfo {author} {\bibfnamefont {M.}~\bibnamefont
  {Fishman}}, \bibinfo {author} {\bibfnamefont {S.~R.}\ \bibnamefont {White}},\
  and\ \bibinfo {author} {\bibfnamefont {E.~M.}\ \bibnamefont {Stoudenmire}},\
  }\href {https://doi.org/10.21468/SciPostPhysCodeb.4-r0.3} {\bibfield
  {journal} {\bibinfo  {journal} {SciPost Phys. Codebases}\ ,\ \bibinfo {pages}
  {4}} (\bibinfo {year} {2022}{\natexlab{b}})}\BibitemShut {NoStop}%
\bibitem [{\citenamefont {Cataldi}(2024)}]{Cataldi2024}%
  \BibitemOpen
  \bibfield  {author} {\bibinfo {author} {\bibfnamefont {G.}~\bibnamefont
  {Cataldi}},\ }\href {https://doi.org/10.5281/ZENODO.11145318} {\bibinfo
  {title} {ed-lgt}},\ \bibinfo {howpublished} {Code for Exact Diagonalization
  of Quantum Many-Body Hamiltonians and Lattice Gauge Theories.} (\bibinfo
  {year} {2024})\BibitemShut {NoStop}%
\bibitem [{\citenamefont {Silvi}\ \emph {et~al.}(2014)\citenamefont {Silvi},
  \citenamefont {Rico}, \citenamefont {Calarco},\ and\ \citenamefont
  {Montangero}}]{Silvi_2014}%
  \BibitemOpen
  \bibfield  {author} {\bibinfo {author} {\bibfnamefont {P.}~\bibnamefont
  {Silvi}}, \bibinfo {author} {\bibfnamefont {E.}~\bibnamefont {Rico}},
  \bibinfo {author} {\bibfnamefont {T.}~\bibnamefont {Calarco}},\ and\ \bibinfo
  {author} {\bibfnamefont {S.}~\bibnamefont {Montangero}},\ }\href
  {https://doi.org/10.1088/1367-2630/16/10/103015} {\bibfield  {journal}
  {\bibinfo  {journal} {New Journal of Physics}\ }\textbf {\bibinfo {volume}
  {16}},\ \bibinfo {pages} {103015} (\bibinfo {year} {2014})}\BibitemShut
  {NoStop}%
\bibitem [{\citenamefont {Gebert}\ \emph {et~al.}(2014)\citenamefont {Gebert},
  \citenamefont {K{\"u}ttler},\ and\ \citenamefont
  {M{\"u}ller}}]{gebert2014anderson}%
  \BibitemOpen
  \bibfield  {author} {\bibinfo {author} {\bibfnamefont {M.}~\bibnamefont
  {Gebert}}, \bibinfo {author} {\bibfnamefont {H.}~\bibnamefont
  {K{\"u}ttler}},\ and\ \bibinfo {author} {\bibfnamefont {P.}~\bibnamefont
  {M{\"u}ller}},\ }\href
  {https://doi.org/https://link.springer.com/article/10.1007/s00220-014-1914-3#citeas}
  {\bibfield  {journal} {\bibinfo  {journal} {Communications in Mathematical
  Physics}\ }\textbf {\bibinfo {volume} {329}},\ \bibinfo {pages} {979}
  (\bibinfo {year} {2014})}\BibitemShut {NoStop}%
\bibitem [{\citenamefont {Zhou}\ \emph {et~al.}(2008)\citenamefont {Zhou},
  \citenamefont {Or\'us},\ and\ \citenamefont
  {Vidal}}]{PhysRevLett.100.080601}%
  \BibitemOpen
  \bibfield  {author} {\bibinfo {author} {\bibfnamefont {H.-Q.}\ \bibnamefont
  {Zhou}}, \bibinfo {author} {\bibfnamefont {R.}~\bibnamefont {Or\'us}},\ and\
  \bibinfo {author} {\bibfnamefont {G.}~\bibnamefont {Vidal}},\ }\href
  {https://doi.org/10.1103/PhysRevLett.100.080601} {\bibfield  {journal}
  {\bibinfo  {journal} {Phys. Rev. Lett.}\ }\textbf {\bibinfo {volume} {100}},\
  \bibinfo {pages} {080601} (\bibinfo {year} {2008})}\BibitemShut {NoStop}%
\bibitem [{\citenamefont {{Fannes}}\ \emph {et~al.}(1992)\citenamefont
  {{Fannes}}, \citenamefont {{Nachtergaele}},\ and\ \citenamefont
  {{Werner}}}]{1992CMaPh.144..443F}%
  \BibitemOpen
  \bibfield  {author} {\bibinfo {author} {\bibfnamefont {M.}~\bibnamefont
  {{Fannes}}}, \bibinfo {author} {\bibfnamefont {B.}~\bibnamefont
  {{Nachtergaele}}},\ and\ \bibinfo {author} {\bibfnamefont {R.~F.}\
  \bibnamefont {{Werner}}},\ }\href {https://doi.org/10.1007/BF02099178}
  {\bibfield  {journal} {\bibinfo  {journal} {Communications in Mathematical
  Physics}\ }\textbf {\bibinfo {volume} {144}},\ \bibinfo {pages} {443}
  (\bibinfo {year} {1992})}\BibitemShut {NoStop}%
\bibitem [{\citenamefont {Klumper}\ \emph {et~al.}(1991)\citenamefont
  {Klumper}, \citenamefont {Schadschneider},\ and\ \citenamefont
  {Zittartz}}]{A_Klumper_1991}%
  \BibitemOpen
  \bibfield  {author} {\bibinfo {author} {\bibfnamefont {A.}~\bibnamefont
  {Klumper}}, \bibinfo {author} {\bibfnamefont {A.}~\bibnamefont
  {Schadschneider}},\ and\ \bibinfo {author} {\bibfnamefont {J.}~\bibnamefont
  {Zittartz}},\ }\href {https://doi.org/10.1088/0305-4470/24/16/012} {\bibfield
   {journal} {\bibinfo  {journal} {Journal of Physics A: Mathematical and
  General}\ }\textbf {\bibinfo {volume} {24}},\ \bibinfo {pages} {L955}
  (\bibinfo {year} {1991})}\BibitemShut {NoStop}%
\bibitem [{\citenamefont {Klümper}\ \emph {et~al.}(1993)\citenamefont
  {Klümper}, \citenamefont {Schadschneider},\ and\ \citenamefont
  {Zittartz}}]{A_Klumper_1993}%
  \BibitemOpen
  \bibfield  {author} {\bibinfo {author} {\bibfnamefont {A.}~\bibnamefont
  {Klümper}}, \bibinfo {author} {\bibfnamefont {A.}~\bibnamefont
  {Schadschneider}},\ and\ \bibinfo {author} {\bibfnamefont {J.}~\bibnamefont
  {Zittartz}},\ }\href {https://doi.org/10.1209/0295-5075/24/4/010} {\bibfield
  {journal} {\bibinfo  {journal} {Europhysics Letters}\ }\textbf {\bibinfo
  {volume} {24}},\ \bibinfo {pages} {293} (\bibinfo {year} {1993})}\BibitemShut
  {NoStop}%
\bibitem [{\citenamefont {Vidal}(2003)}]{vidal2003efficient}%
  \BibitemOpen
  \bibfield  {author} {\bibinfo {author} {\bibfnamefont {G.}~\bibnamefont
  {Vidal}},\ }\href@noop {} {\bibfield  {journal} {\bibinfo  {journal}
  {Physical review letters}\ }\textbf {\bibinfo {volume} {91}},\ \bibinfo
  {pages} {147902} (\bibinfo {year} {2003})}\BibitemShut {NoStop}%
\bibitem [{\citenamefont {Orús}(2014)}]{ORUS2014117}%
  \BibitemOpen
  \bibfield  {author} {\bibinfo {author} {\bibfnamefont {R.}~\bibnamefont
  {Orús}},\ }\href {https://doi.org/https://doi.org/10.1016/j.aop.2014.06.013}
  {\bibfield  {journal} {\bibinfo  {journal} {Annals of Physics}\ }\textbf
  {\bibinfo {volume} {349}},\ \bibinfo {pages} {117} (\bibinfo {year}
  {2014})}\BibitemShut {NoStop}%
\bibitem [{\citenamefont {Eisert}\ \emph {et~al.}(2010)\citenamefont {Eisert},
  \citenamefont {Cramer},\ and\ \citenamefont {Plenio}}]{RevModPhys.82.277}%
  \BibitemOpen
  \bibfield  {author} {\bibinfo {author} {\bibfnamefont {J.}~\bibnamefont
  {Eisert}}, \bibinfo {author} {\bibfnamefont {M.}~\bibnamefont {Cramer}},\
  and\ \bibinfo {author} {\bibfnamefont {M.~B.}\ \bibnamefont {Plenio}},\
  }\href {https://doi.org/10.1103/RevModPhys.82.277} {\bibfield  {journal}
  {\bibinfo  {journal} {Rev. Mod. Phys.}\ }\textbf {\bibinfo {volume} {82}},\
  \bibinfo {pages} {277} (\bibinfo {year} {2010})}\BibitemShut {NoStop}%
\bibitem [{\citenamefont {Verstraete}\ \emph {et~al.}(2008)\citenamefont
  {Verstraete}, \citenamefont {Murg},\ and\ \citenamefont
  {Cirac}}]{verstraete2008matrix}%
  \BibitemOpen
  \bibfield  {author} {\bibinfo {author} {\bibfnamefont {F.}~\bibnamefont
  {Verstraete}}, \bibinfo {author} {\bibfnamefont {V.}~\bibnamefont {Murg}},\
  and\ \bibinfo {author} {\bibfnamefont {J.~I.}\ \bibnamefont {Cirac}},\
  }\href@noop {} {\bibfield  {journal} {\bibinfo  {journal} {Advances in
  physics}\ }\textbf {\bibinfo {volume} {57}},\ \bibinfo {pages} {143}
  (\bibinfo {year} {2008})}\BibitemShut {NoStop}%
\bibitem [{\citenamefont {Schollw{\"o}ck}(2011)}]{schollwock2011density}%
  \BibitemOpen
  \bibfield  {author} {\bibinfo {author} {\bibfnamefont {U.}~\bibnamefont
  {Schollw{\"o}ck}},\ }\href@noop {} {\bibfield  {journal} {\bibinfo  {journal}
  {Annals of physics}\ }\textbf {\bibinfo {volume} {326}},\ \bibinfo {pages}
  {96} (\bibinfo {year} {2011})}\BibitemShut {NoStop}%
\bibitem [{\citenamefont {Vidal}(2004)}]{PhysRevLett.93.040502}%
  \BibitemOpen
  \bibfield  {author} {\bibinfo {author} {\bibfnamefont {G.}~\bibnamefont
  {Vidal}},\ }\href {https://doi.org/10.1103/PhysRevLett.93.040502} {\bibfield
  {journal} {\bibinfo  {journal} {Phys. Rev. Lett.}\ }\textbf {\bibinfo
  {volume} {93}},\ \bibinfo {pages} {040502} (\bibinfo {year}
  {2004})}\BibitemShut {NoStop}%
\bibitem [{\citenamefont {Haegeman}\ \emph {et~al.}(2011)\citenamefont
  {Haegeman}, \citenamefont {Cirac}, \citenamefont {Osborne}, \citenamefont
  {Pi\ifmmode~\check{z}\else \v{z}\fi{}orn}, \citenamefont {Verschelde},\ and\
  \citenamefont {Verstraete}}]{PhysRevLett.107.070601}%
  \BibitemOpen
  \bibfield  {author} {\bibinfo {author} {\bibfnamefont {J.}~\bibnamefont
  {Haegeman}}, \bibinfo {author} {\bibfnamefont {J.~I.}\ \bibnamefont {Cirac}},
  \bibinfo {author} {\bibfnamefont {T.~J.}\ \bibnamefont {Osborne}}, \bibinfo
  {author} {\bibfnamefont {I.}~\bibnamefont {Pi\ifmmode~\check{z}\else
  \v{z}\fi{}orn}}, \bibinfo {author} {\bibfnamefont {H.}~\bibnamefont
  {Verschelde}},\ and\ \bibinfo {author} {\bibfnamefont {F.}~\bibnamefont
  {Verstraete}},\ }\href {https://doi.org/10.1103/PhysRevLett.107.070601}
  {\bibfield  {journal} {\bibinfo  {journal} {Phys. Rev. Lett.}\ }\textbf
  {\bibinfo {volume} {107}},\ \bibinfo {pages} {070601} (\bibinfo {year}
  {2011})}\BibitemShut {NoStop}%
\bibitem [{\citenamefont {Paeckel}\ \emph {et~al.}(2019)\citenamefont
  {Paeckel}, \citenamefont {Köhler}, \citenamefont {Swoboda}, \citenamefont
  {Manmana}, \citenamefont {Schollwöck},\ and\ \citenamefont
  {Hubig}}]{PAECKEL2019167998}%
  \BibitemOpen
  \bibfield  {author} {\bibinfo {author} {\bibfnamefont {S.}~\bibnamefont
  {Paeckel}}, \bibinfo {author} {\bibfnamefont {T.}~\bibnamefont {Köhler}},
  \bibinfo {author} {\bibfnamefont {A.}~\bibnamefont {Swoboda}}, \bibinfo
  {author} {\bibfnamefont {S.~R.}\ \bibnamefont {Manmana}}, \bibinfo {author}
  {\bibfnamefont {U.}~\bibnamefont {Schollwöck}},\ and\ \bibinfo {author}
  {\bibfnamefont {C.}~\bibnamefont {Hubig}},\ }\href
  {https://doi.org/https://doi.org/10.1016/j.aop.2019.167998} {\bibfield
  {journal} {\bibinfo  {journal} {Annals of Physics}\ }\textbf {\bibinfo
  {volume} {411}},\ \bibinfo {pages} {167998} (\bibinfo {year}
  {2019})}\BibitemShut {NoStop}%
\bibitem [{\citenamefont {Sandvik}(2010)}]{Sandvik2010}%
  \BibitemOpen
  \bibfield  {author} {\bibinfo {author} {\bibfnamefont {A.~W.}\ \bibnamefont
  {Sandvik}},\ }\href {https://doi.org/10.1063/1.3518900} {\bibfield  {journal}
  {\bibinfo  {journal} {AIP Conference Proceedings}\ }\textbf {\bibinfo
  {volume} {1297}},\ \bibinfo {pages} {135} (\bibinfo {year}
  {2010})}\BibitemShut {NoStop}%
\end{thebibliography}%

\end{document}